\documentclass[reprint,twocolumn,amsmath,amssymb,superscriptaddress]{revtex4-2}

\usepackage{graphicx, graphpap}
\usepackage[dvipsnames]{xcolor}
\usepackage[caption=false,subrefformat=parens,labelformat=parens]{subfig}
\usepackage{amsmath}
\usepackage{overpic}
\usepackage{physics}
\usepackage{enumitem}
\usepackage{comment}
\usepackage{amssymb}
\usepackage{amsthm}
\usepackage{pstricks}
\usepackage{float}
\usepackage{multirow}
\usepackage{hyperref}
\usepackage{graphicx}
\usepackage[T1]{fontenc}
\usepackage{framed} 
\usepackage{bm}
\usepackage{bbm}

\definecolor{shadecolor}{gray}{0.9}
\definecolor{ECCgreen}{RGB}{0, 153, 0}
\definecolor{EditPurple}{RGB}{160, 32, 240}

\renewcommand{\Tr}[1]{\mathrm{Tr}\left[#1\right]}

\usepackage{placeins}

\newtheorem{theorem}{Theorem}

\begin{document}
	
\title{Composable Finite-Size Security of High-Dimensional Quantum Key Distribution Protocols}

\author{Florian Kanitschar}
\email{florian.kanitschar@outlook.com}
\affiliation{Vienna Center for Quantum Science and Technology (VCQ), Atominstitut, Technische Universität Wien, Stadionallee 2, 1020 Vienna, Austria}
\affiliation{AIT  Austrian  Institute  of  Technology,  Center  for  Digital  Safety\&Security,  Giefinggasse  4,  1210  Vienna, Austria}
\author{Marcus Huber}
\affiliation{Vienna Center for Quantum Science and Technology (VCQ), Atominstitut, Technische Universität Wien, Stadionallee 2, 1020 Vienna, Austria}
 \affiliation{Institute for Quantum Optics and Quantum Information (IQOQI),
 Austrian Academy of Sciences, Boltzmanngasse 3, 1090 Vienna, Austria}
\date{\today}


\begin{abstract}
Practical implementations of Quantum Key Distribution (QKD) extending beyond urban areas commonly use satellite links. However, the transmission of quantum states through the Earth's atmosphere is highly susceptible to noise, restricting its application primarily to nighttime. High-dimensional (HD) QKD offers a promising solution to this limitation by employing high-dimensionally entangled quantum states. Although experimental platforms for HD QKD exist, previous security analyses have been limited to the asymptotic regime and have either relied on impractical measurements or employed computationally demanding convex optimization tasks restricting the security analysis to low dimensions. In this work, we bridge this gap by presenting a composable finite-size security proof against both collective and coherent attacks for a general HD QKD protocol that uses only experimentally accessible measurements. In addition to the conventional, yet impractical `one-shot' key rates, we provide a practical variable-length security argument that yields significantly higher expected key rates. This approach is particularly crucial under rapidly changing and turbulent atmospheric conditions, as encountered in free-space and satellite-based QKD platforms.
\end{abstract}
	
\maketitle


\twocolumngrid
\section{Introduction}\label{sec:Introduction}
Quantum Key Distribution \cite{Bennett_Brassard_1984, Ekert_1991} enables two remote parties, commonly called Alice and Bob, to establish cryptographically secure keys, without relying on assumptions on the hardness of mathematical problems. Over recent decades, significant advancements in both experimental techniques and security proofs have evolved QKD from an interesting scientific concept toward one of the most developed quantum technologies \cite{Scarani_2009, Pirandola_2020, Usenko_2025}, and it is now on the edge of practical use for both private and governmental applications. Implementations range from fiber links \cite{Hiskett_2006, SECOQC_2009, Rosenberg_2007, Peng_2007, Boaron_2018} for urban areas to free-space \cite{Manderbach_2007, Krzic_2023} and satellite-based links \cite{Qubesat_2018, Micius_2022, QEYSSAT_2023, Sidhu_2023, Orsucci_2024} that span countries and even continents. Although the transmission of quantum signals through the atmosphere offers a more favorable loss profile compared to fiber-based systems, it also suffers from high noise sensitivity, which essentially restricts operation to nighttime hours. 

High-dimensional quantum states \cite{Kwiat_1997, Barreiro_2005, Martin_2017} have proven to enhance noise tolerance in entanglement distribution tasks \cite{Ecker_2019}, making high-dimensional quantum key distribution (HD QKD) a promising candidate for improving noise robustness in cryptographic tasks. Nonetheless, the security analysis of practically implementable HD QKD protocols is so far rather uncharted. Previous works have primarily focused on asymptotic security, employing either analytical methods \cite{Sheridan_2010,Ferenczi_2012, Doda_2021, Wyderka_2025, Thomas_2025} that leverage symmetry arguments or numerical approaches \cite{BKPH_2023} involving convex optimization \cite{Coles_2016, Winick_2018, Araujo_2022}. Analytical arguments often require $2$ or even $d+1$ mutually unbiased bases (MUB) measurements, which are impractical, particularly for large dimensions. Numerical methods avoid this but require solving very high-dimensional convex optimization problems, which are computationally demanding and very memory-intensive, rendering them infeasible even for low to medium dimensions \cite{BKPH_2023}. Our recent work in Ref. \cite{Kanitschar_2024a} has closed this gap between experimentally available setups \cite{Bavaresco_2018, Hu_2020, Schneeloch_2019, Ponce_2022, Maxwell_2022} and asymptotic security, by providing an efficient semianalytic method to evaluate key rates for practical HD QKD setups, relying solely on practically implementable measurements. Yet, the security of high-dimensional QKD protocols in the finite-size regime remains an open question. 

In this study, we address this issue by presenting a full, composable finite-size security proof for general high-dimensional QKD protocols. We begin by outlining the general HD QKD protocol in Section \ref{sec:ProtocolDescription}. Section \ref{sec:SecurityArgument} details our statistical testing procedure and establishes composable finite-size security against i.i.d. collective attacks. We then apply the postselection technique \cite{Christandl_2009, Nahar_2024} to extend security to coherent attacks. These secure key rates refer to the so-called fixed-length scenario, which is commonly reported in the literature. However, in practical settings, in particular for atmospheric channels, the quantum channel's instability complicates the pre-definition of accepted statistics before the execution of the protocol, as required by the security argument. This leads to a high abort rate of the protocol and low effective key rates. To address this, in Section \ref{sec:VarLengthSecurity}, we give a variable-length security argument for a variable-length version of our protocol \cite{Tupkary_2024}, again proving composable finite-size variable-length security against both i.i.d. collective and coherent attacks. Finally, in Section \ref{sec:ProofMethod}, we apply our recent method \cite{Kanitschar_2024a} and present both variable-length and fixed-length key rates in Section \ref{sec:Results}.

\section{Protocol}\label{sec:ProtocolDescription}
We analyze a general high-dimensional, entanglement-based QKD protocol of dimension $d$, which consists of the following steps;

\textbf{\textit{1.) State Generation.}} A photon source distributes entangled quantum states $\rho_{AB}$ to Alice and Bob. 

\textbf{\textit{2.) Measurement.}} Alice and Bob randomly and independently decide to measure either in their computational bases, represented by POVMs $\{A_1^x\}_{x=0}^{d-1}, \{B_1^y\}_{y=0}^{d-1}$, or in one of the test bases, represented by POVMs $\{A_b^x\}_{x=0}^{d-1}, \{B_b^y\}_{y=0}^{d-1}$ for $b \in \{2, ..., n_{\mathrm{bases}}\}$, where $n_{\mathrm{bases}}$ denotes the total number of bases used in the protocol, and record their outcomes in their respective private registers $\bar{A}$ and $\bar{B}$. Practically, this can be implemented either passively via beamsplitters or using active switching with random numbers. We denote the probability that basis $i$ is chosen by $p_i\geq 0$, and $d$ is the dimension of the protocol. 

Steps 1.) and 2.) are repeated $N$ times. Once the measurements are completed, Alice may announce a random permutation, which both parties then apply to their data. 

\textbf{\textit{3.) Public Announcements \& Sifting.}} Alice and Bob use the classical authenticated channel to communicate their measurement choice to each other and may discard certain results. Therefore, Bob announces the time stamps of his single clicks, and Alice checks which of those singles lie in her pre-defined coincidence window. Then, Alice announces her random measurement setting for those rounds, where she recorded a single in the corresponding window. Other rounds are discarded. Their announcements are stored in their public registers $\tilde{A}$ and $\tilde{B}$. 

\textbf{\textit{4.) Acceptance Testing.}} The communicating parties disclose a randomly chosen subset consisting of $k_T$ of their measurement results over the public channel and perform a so-called acceptance test (see Theorem \ref{Thm:AcceptanceTest}). This test helps them to define a set $\mathcal{S}^{\mathrm{AT}}$. They may also announce that they will perform subspace postselection. 

\textbf{\textit{5.) Key Map}} Based on their announcements and their measurement results one party performs a key map $g$, which formalizes the process of assigning a logical bit value $\{0, ..., d-1\}$ to each of the remaining $N-k_T$ rounds as a function of the publicly known announcements and the private measurement data. This step also may allow discarding certain rounds (by assigning the symbol $\perp$ or, more efficiently, by directly shortening the string \cite{Tupkary_2024}). In case Alice performs the key map, this is called direct reconciliation; in case it is done by Bob, it is called reverse reconciliation. 

\textbf{\textit{6.) Error-Correction \& Error-Verification}} On the remaining $N-k_T$ rounds, Alice and Bob perform error-correction to reconcile their raw keys $X$ and $Y$, which leaks some information to Eve. This is followed by error-verification, where Alice chooses a two-universal hash function that hashes to $\left\lceil \log_2\left(\frac{1}{\epsilon_{\mathrm{EV}}}\right) \right\rceil$ bits and sends them to Bob, who performs the same hash function and compares them. 

\textbf{\textit{7.) Privacy Amplification.}} Finally, they perform privacy amplification, where they map their reconciled key to a shorter one, aiming to decouple Eve from their shared randomness.

\section{Security Argument}\label{sec:SecurityArgument}
In this work, we aim to demonstrate the composable security of a generic high-dimensional QKD protocol against coherent attacks in the finite-size regime. This task, however, presents several challenges compared to the asymptotic analysis carried out in Ref. \cite{Kanitschar_2024a}. Although in the asymptotic regime, the law of large numbers justifies assuming that our measurement outcomes are known with certainty, in the finite-size regime, we need to thoroughly design and perform an acceptance test. The acceptance test carefully compares the observations with the expected channel behaviour. If the test passes, the analysis proceeds on the set of all states that are $\epsilon_{\mathrm{AT}}$-likely to be compatible with our observations. Instead of considering a plethora of attack strategies that Eve could follow (a nearly impossible task), this allows us to base the security argument solely on the observations made by the communicating parties. However, the direct use of the measured statistics, as it is common in discrete-variable QKD, for HD-QKD performed in practical measurement setups requires numerical security arguments. As identified in Ref. \cite{BKPH_2023}, these methods suffer from serious limitations due to complexity issues, which motivated the development of a new, semi-analytic security argument for practical HD-QKD setups in Ref. \cite{Kanitschar_2024a}. This work avoids the use of full statistics, allowing for direct solutions and bypassing highly demanding numerical convex optimization routines. However, it also means that we have to construct the acceptance set based on suitable coarse-grained statistics. Additionally, we need to take into account that all subroutines used for the QKD protocol may fail with a certain probability, such as error verification failing to align the keys between Alice and Bob or privacy amplification failing to decouple Eve from the final key. Each of these contributions has to be taken into account carefully in a composable security argument. Proving composable finite-size security against coherent attacks is inherently challenging, as most available arguments rely on an assumed i.i.d. structure of the underlying quantum states. Consequently, a lift from collective i.i.d. to coherent attacks is required.

We address these issues as follows. Initially, we consider collective i.i.d. attacks, where Eve prepares a fresh ancilla state that interacts identically with each protocol round. She then may store this ancilla state in her quantum memory and wait until Alice and Bob have finished executing their QKD protocol, including their announcements, before measuring her quantum memory. Mathematically, this implies that different rounds are uncorrelated and that Alice's and Bob's shared state has tensor-product structure $\rho = \rho_{AB}^{\otimes N}$. We then adapt an argument by Refs. \cite{George_2021, Kanitschar_2023} to our needs and formulate an acceptance test with non-unique acceptance criteria. Based on this test, we prove composable security within Renner's $\epsilon$-security framework \cite{Renner_2005}. Finally, we apply the postselection technique \cite{Christandl_2009, Nahar_2024} to lift our collective i.i.d. key rates to secure key rates against coherent attacks, thereby completing the proof. 

\subsection{Security against i.i.d. collective attacks}
The first goal of our security argument is to prove the security of the considered protocol against coherent attacks. We aim to tackle this task in two steps: We first prove security against i.i.d. collective attacks, and we then lift our proof to coherent attacks. For our collective attacks argument, we follow the $\epsilon$-security framework by Renner~\cite{Renner_2005}. Thus, for now, let us assume i.i.d. collective attacks, i.e., Alice's and Bob's shared quantum state after $N$ rounds of transmission has tensor-product structure $\rho_{AB} = \tau_{AB}^{\otimes N}$. Let $\rho_{ABE} = \tau_{ABE}^{\otimes N}$ be a purification of this state, held by Eve. Alice and Bob then randomly select $k_T$ of those rounds for statistical testing. 

Formally, this means we need to construct a set of $\epsilon_{\mathrm{AT}}$-filtered states \cite{Renner_2005, George_2021}, which is a set $\left(\mathcal{S}^{\mathrm{AT}}\right)^{\mathsf{c}}$ of quantum states $\sigma$ leading to an accepted statistics with probability less than $\epsilon_{\mathrm{AT}}$, $\mathrm{Pr}\left[\mathrm{accept} | \sigma\right] \leq \epsilon_{\mathrm{AT}}$. In other words, our goal is to define a set $\mathcal{S}^{\mathrm{AT}}$ such that all states in its complement $\left(\mathcal{S}^{\mathrm{AT}}\right)^{\mathsf{c}}$ could have generated the accepted statistics only with probability less than $\epsilon_{\mathrm{AT}}$. Intuitively, this allows us to split the set of all quantum states into two subsets: the set of states that are very unlikely to have produced the observed statistics, and the set of states that have potentially caused our observations. We formalise this notion in the following theorem, which is an adapted version of Ref. \cite[Theorem 3]{Kanitschar_2023}.

\begin{theorem}[\textbf{Acceptance Test \cite{Kanitschar_2023}}]\label{Thm:AcceptanceTest}
Let $\Theta$ be the set of Alice's and Bob's observables. Let $\mathbf{r} \in \mathbb{R}^{|\Theta|}$ and $\mathbf{t} \in \mathbb{R}^{|\Theta|}_{\geq 0}$, where $|\Theta|$ denotes the cardinality of $\Theta$ and choose $\epsilon_{\mathrm{AT}} > 0$. Define the set of accepted statistics as
\begin{equation}\label{eq:accepted-observations} \mathcal{O} := \{ \mathbf{v} \in \mathbb{R}^{|\Theta|} : \forall X \in \Theta, |v_{X} - r_{X}| \leq t_{X} \} \ ,
\end{equation}
and the corresponding acceptance set as
\begin{equation}\label{eq:ATset}
\begin{aligned}
    &\mathcal{S}^{\mathrm{AT}}:=  \left\{ \sigma \in \mathcal{D}(\mathcal{H}_A \otimes \mathcal{H}_B \otimes \mathcal{H}_E):\right. \\
    & \hspace{18mm} \forall X \in \Theta, |\Tr{\sigma X} - r_{X}| \leq \mu_{X} + t_{X} \},
\end{aligned}
\end{equation}
where $r_{X}$ is the $X$-th element of the vector $\mathbf{r}$ and likewise for $t_{X}$. For every $X \in \Theta$, let \begin{equation*}
    \mu_X := \sqrt{\frac{2x^2}{m_{X}} \ln\left( \frac{2}{\epsilon_{\mathrm{AT}}} \right)}
\end{equation*}
where $x := \|X\|_{\infty}$ and $m_{X}$ is the number of tests for the observable $X$. If $\rho \not \in \mathcal{S}^{\mathrm{AT}}$, then the probability of accepting the statistics generated by the i.i.d. measurements of $\rho^{\otimes n}$ is bounded above by $\epsilon_{\mathrm{AT}}$. That is, the complement of $\mathcal{S}^{\mathrm{AT}}$ consists of all $\epsilon_{\mathrm{AT}}$-filtered states.
\end{theorem}
Let us denote the event of the acceptance test passing by $\mathcal{P}$. Note that we never actually know which state Alice and Bob received, but only use the measurement results and the testing procedure to decide how to proceed with the protocol. Thus, we face four different cases
\begin{enumerate}
    \item[a)] $ \left(\tau_{AB} \in \mathcal{S}^{\mathrm{AT}} \right) \land \mathcal{P}_{\mathrm{AT}}$  -- the test passes for an input state in $\mathcal{S}^{\mathrm{AT}}$,
    \item[b)] $ \left(\tau_{AB} \in \mathcal{S}^{\mathrm{AT}} \right) \land \mathcal{P}_{\mathrm{AT}}^{\mathsf{c}}$  -- the test fails for an input state in $\mathcal{S}^{\mathrm{AT}}$,
    \item[c)] $ \left(\tau_{AB} \notin \mathcal{S}^{\mathrm{AT}} \right) \land \mathcal{P}_{\mathrm{AT}}$  -- the test passes for an input state not in $\mathcal{S}^{\mathrm{AT}}$,
    \item[d)] $ \left(\tau_{AB} \notin \mathcal{S}^{\mathrm{AT}} \right) \land \mathcal{P}_{\mathrm{AT}}^{\mathsf{c}}$  -- the test fails for an input state not in $\mathcal{S}^{\mathrm{AT}}$.
\end{enumerate}
While cases a) and d) correspond to the desired behavior of the test, note that in case b), although the protocol erroneously aborts, it is nevertheless secure. By design of the acceptance testing theorem, the probability that c) occurs is bounded by $\epsilon_{\mathrm{AT}}$. Additionally, we denote the event that the error-verification routine passes by $\mathcal{P}_{\mathrm{EV}}$ and abbreviate the event of both passing by $\mathcal{P}$.

Composable security builds upon the following Gedankenexperiment \cite{Portmann_2022}. The adversary is given the output of both the real and an idealized version of the QKD protocol and aims to distinguish between both protocols without having access to Alice's and Bob's private data. The ideal output is uniformly distributed and completely decoupled from Eve, $\rho_{\mathrm{ideal}} := \pi_{K_A K_B} \otimes \rho_E$, where $\pi_{K_A K_B} := \frac{1}{|\mathcal{K}|} \sum_{k \in \mathcal{K}} \ketbra{k} \otimes \ketbra{k}$. The output of the real protocol shall be denoted by $\rho_{K_A K_B E} := \Phi(\rho_{ABE})$, where $\Phi$ denotes the map representing the action of the protocol (which will be detailed on in Section \ref{sec:PPMap}). Then, we aim to bound the distinguishing advantage of an adversary, 
\begin{equation}
    \frac{1}{2} \left|\left| \rho_{K_A K_B E} - \pi_{K_AK_B} \otimes \rho_E  \right|\right|_1 \leq \epsilon.
\end{equation}
Since the protocol aborts and is trivially secure when the acceptance test fails (cases b) and d) ) there is no distinguishing advantage for Eve in that case. Thus, we can focus on the cases where the test passes,
\begin{align*}
    &\frac{1}{2} \left|\left| \rho_{K_A K_B E} - \pi_{K_AK_B} \otimes \rho_E  \right|\right|_1 \\
    =& 0 +\frac{1}{2} \left|\left| \rho_{K_A K_B E \land \mathcal{P}} - \pi_{K_AK_B} \otimes \rho_{E \land \mathcal{P}}  \right|\right|_1 \\
    \leq& \frac{1}{2} \left|\left| \rho_{K_A K_B E \land \mathcal{P}} - \rho_{K_A K_A E \land \mathcal{P}}\right| \right|_1\\
    &+ \frac{1}{2} \left|\left| \rho_{K_A K_A E\land \mathcal{P}} - \pi_{K_AK_B} \otimes \rho_{E \land \mathcal{P}}  \right|\right|_1 \\
    \leq & \epsilon_{\mathrm{EV}} + \frac{1}{2} \left|\left| \rho_{K_A K_A E \land \mathcal{P}} - \pi_{K_AK_B} \otimes \rho_{E \land \mathcal{P}}  \right|\right|_1,
\end{align*}
where we used $\mathrm{Pr}\left[\mathcal{P}\right] \sigma_{A|\mathcal{P}} =: \sigma_{A \land \mathcal{P}}$.
For the first inequality, we applied the triangle inequality, and for the second inequality, we used that the first term is bounded by $\epsilon_{\mathrm{EV}}$. To ease notation, we now drop Bob's key register, as it is equal to Alice's. For the second term, we use the fact that the remaining cases a) and c) are mutually exclusive, which leads to the upper bound
\begin{align*}
    &\frac{1}{2} \left|\left| \rho_{K_A K_B E\land \mathcal{P}} - \pi_{K_AK_B} \otimes \rho_{E \land \mathcal{P}}  \right|\right|_1 \\
    &\leq \max \left\{ \mathrm{Pr}\left[\tau_{AB} \in \mathcal{S}^{\mathrm{AT}}\right] \frac{1}{2} \left|\left| \rho_{K_A E \land \mathcal{P}} - \pi_{K_A} \otimes \rho_{E \land \mathcal{P}}  \right|\right|_1, \right. \\
    &~~~~~~~~~\left. \mathrm{Pr}\left[\tau_{AB} \notin \mathcal{S}^{\mathrm{AT}}\right] \frac{1}{2} \left|\left| \rho_{K_A E\land \mathcal{P}} - \pi_{K_A} \otimes \rho_{E \land \mathcal{P}}  \right|\right|_1 \right\}.
\end{align*}
For a state $\tau_{AB} \notin \mathcal{S}^{\mathrm{AT}}$ the distinguishability given that the test passes can be upper bounded by the probability of the test passing on a state that is not in the acceptance set, which in turn is upper bounded by $\epsilon_{\mathrm{AT}}$,
\begin{align*}
    &\mathrm{Pr}\left[\tau_{AB} \notin \mathcal{S}^{\mathrm{AT}}\right] \frac{1}{2} \left|\left| \rho_{K_A E \land \mathcal{P}} - \pi_{K_A} \otimes \rho_{E \land \mathcal{P}}  \right|\right|_1\\
    &\leq \mathrm{Pr}\left[ \mathcal{P}_{\mathrm{AT}}~|~\tau_{AB} \notin \mathcal{S}^{\mathrm{AT}}\right] \leq \epsilon_{\mathrm{AT}}.
\end{align*}

While we have now successfully treated cases b) - d), we are left with case a), where Alice's and Bob's shared key is not fully private with respect to Eve. To decouple their key from Eve, the protocol requires Alice and Bob to perform privacy amplification. The leftover hashing lemma (LHL) \cite[Corollary 5.6.1]{Renner_2005} relates the length of the hashed key to the security parameter,
\begin{align*}
 &\frac{1}{2} \left|\left| \rho_{K_A E\land\mathcal{P}} - \pi_{K_A} \otimes \rho_{E \land \mathcal{P}}  \right|\right|_1 \\
 &\leq 2 \tilde{\epsilon} + 2^{-\frac{1}{2} \left( H_{\mathrm{min}}^{\tilde{\epsilon}}(X|EC)_{\Phi^{(n)}(\rho_{ABE})\land \mathcal{P} }  - \ell\right)} \leq \epsilon_{\mathrm{sec}},
\end{align*}
where $C$ denotes the transcript of the information reconciliation phase. Thus, provided that the length of the hashed key satisfies
\begin{align*}
    \ell \leq H_{\mathrm{min}}^{\tilde{\epsilon}}(X|EC)_{\Phi^{(n)}(\rho_{ABE})\land \mathcal{P}} + 2 \log_2\left(\epsilon_{\mathrm{sec}} - 2 \tilde{\epsilon} \right),
\end{align*}
the key is secret with secrecy parameter $\epsilon_{\mathrm{sec}}$. Next, we use \cite[Lemma 10]{Tomamichel_2017} $H_{\mathrm{min}}^{\tilde{\epsilon}}(X|EC)_{\sigma \land \mathcal{P}} \geq H_{\mathrm{min}}^{\tilde{\epsilon}}(X|EC)_{\sigma}$ as well as Lemma 6.4.1 in \cite{Renner_2005} to remove the information reconciliation transcript from the smooth min-entropy at the cost of introducing a leakage term $\mathrm{leak}_{\mathrm{IR}}$ and arrive at
\begin{align*}
    \ell \leq H_{\mathrm{min}}^{\tilde{\epsilon}}(X|E)_{\Phi^{(n)}(\rho_{ABE})} - \mathrm{leak}_{\mathrm{IR}} + 2 \log_2\left(\epsilon_{\mathrm{sec}} - 2 \tilde{\epsilon} \right).
\end{align*}
Our argument applies to every state in the acceptance set. However, as we are aiming for a reliable bound on the secure key rate, we need to find the worst-case state in this set, which is the state that minimizes the smooth min-entropy. This leads to
\begin{align*}
    \ell \leq \min_{\rho_{ABE} \in \mathcal{S}^{\mathrm{AT}}} H_{\mathrm{min}}^{\tilde{\epsilon}}(X|E)_{\Phi^{(n)}(\rho_{ABE})} &- \mathrm{leak}_{\mathrm{IR}} \\
    &+ 2 \log_2\left(\epsilon_{\mathrm{sec}}- 2 \tilde{\epsilon} \right).
\end{align*}

Let $n:= N - k_T$ denote the number of rounds used for key generation. Since the smooth min-entropy of a quantum state is defined as the maximum of the min-entropy over all quantum states in an $\epsilon$-ball around the considered quantum state, we can lower bound $H_{\mathrm{min}}^{\tilde{\epsilon}}(X|E)_{\Phi(\rho_{ABE})} \geq H_{\mathrm{min}}(X|E)_{\Phi(\rho_{ABE})}$ (i.e., set $\tilde{\epsilon} = 0$). Next, we use the fact that the min-entropy is additive for i.i.d. copies \cite[Table 4.1]{Thomamichel_Thesis} and obtain
\begin{align*}
    \ell \leq&~ n \min_{\tau_{ABE} \in \mathcal{S}^{\mathrm{AT}}} H(X|E)_{\Phi(\tau_{ABE})} - \mathrm{leak}_{\mathrm{IR}} + 2 \log_2\left(\epsilon_{\mathrm{sec}} \right),
\end{align*}
Thus, we choose $\epsilon_{\mathrm{sec}} = \epsilon_{\mathrm{PA}}$ and obtain a valid lower bound on the secure key rate provided that we choose $\ell$ such that
\begin{equation}
\begin{aligned}
      \ell \leq& n \min_{\tau_{ABE} \in \mathcal{S}^{\mathrm{AT}}} H_{\mathrm{min}}(X|E)_{\Phi(\tau_{ABE})} - \mathrm{leak}_{\mathrm{IR}} - 2 \log_2\left(\frac{1}{\epsilon_{\mathrm{PA}}}\right)  
\end{aligned}
\end{equation}
with total security parameter $\epsilon := \epsilon_{\mathrm{EV}} + \max\left\{\epsilon_{\mathrm{AT}}, \epsilon_{\mathrm{PA}}\right\}$. This concludes the security argument for i.i.d. collective attacks. We summarize this in the following theorem.

\begin{theorem}[\textbf{Security statement for collective attacks}]\label{thm:SecurityStatement}
    Let $\epsilon_{\mathrm{EV}}, \epsilon_{\mathrm{PA}}, \epsilon_{\mathrm{AT}} >0$. Let $N$ be the total number of quantum signals exchanged, $k_T$ the number of test rounds, $n:=N-k_T$. The general high-dimensional QKD protocol described in Section \ref{sec:ProtocolDescription} is $\epsilon_{\mathrm{EV}} + \max\left\{\epsilon_{\mathrm{AT}},\epsilon_{\mathrm{PA}} \right\}$-secure against i.i.d. collective attacks, if, in case the protocol does not abort, the secure key length $\ell$ is chosen such that
   \begin{equation}\label{eq:secStatement_eq}
\begin{aligned}
      \ell \leq& n \min_{\tau_{ABE} \in \mathcal{S}^{\mathrm{AT}}} H_{\mathrm{min}}(X|E)_{\Phi(\tau_{ABE})}- \mathrm{leak}_{\mathrm{IR}} - 2 \log_2\left(\frac{1}{\epsilon_{\mathrm{PA}}}\right),  
\end{aligned}
\end{equation}
where $\mathrm{leak}_{\mathrm{IR}}$ denotes the leakage due to the information reconciliation phase and $\mathcal{S}^{\mathrm{AT}}$ is the acceptance set defined by the acceptance test in Theorem \ref{Thm:AcceptanceTest}.
\end{theorem}

Leaving aside the technical details, for pre-fixed subroutine security parameters $\epsilon_{\mathrm{EV}}, \epsilon_{\mathrm{PA}}$ and $\epsilon_{\mathrm{AT}}$ this theorem relates the achievable secure key length $\ell$ under collective attacks to an optimisation problem over the set $\mathcal{S}^{\mathrm{AT}}$, which we defined earlier in the acceptance test, along with correction terms due to error correction and privacy amplification. It turns out that the remaining optimisation problem can now be solved using the method we introduced in our companion paper \cite{Kanitschar_2024a}.

\subsection{Lift to coherent attacks}
It remains to lift the proof to coherent attacks, based on the postselection technique by Refs. \cite{Christandl_2009, Nahar_2024}. We follow the notation used in Ref. \cite{Nahar_2024}. This requires some additional notation. Recall that $\Phi:~ ABE \mapsto K_A K_B E$, where Eve's register $E$ contains all information she gathered during the protocol execution and define $\Phi^{(\ell)} := \mathrm{Tr}_E \circ \mathrm{Tr}_{K_B} \circ \Phi$, a QKD protocol map that produces a key of length $\ell$ upon acceptance. Additionally, let $\Phi^{(\ell)}_{\mathrm{ideal}}$ be the corresponding ideal QKD protocol map. Now, in contrast to the first part of the proof, we now assume a general $\rho_{ABE} = \rho_{A^n B^n E^n}$ without any known structure, so in particular $\rho_{ABE} \neq \tau_{ABE}^{\otimes n}$. After splitting off the correctness part, similar to the argument above, the secrecy statement reads
\begin{align}
    \frac{1}{2}\left|\left| \left(\left( \Phi^{(\ell)} - \Phi_{\mathrm{ideal}}^{(\ell)}\right) \otimes \mathrm{id}_E^n \right) \left(\rho_{A^nB^nE^n} \right) \right|\right|_1 \leq \epsilon_{\mathrm{sec}}.
\end{align}

Since we applied a random permutation to the classical data acquired during the protocol execution, for measurements that act only on single rounds, the considered protocol is permutation invariant such that the map $\Phi^{(\ell)} - \Phi_{\mathrm{ideal}}^{(\ell)}$ is also permutation invariant (for a formal definition of permutation invariance of a map, we refer to \cite{Christandl_2009, Nahar_2024}). For such maps, the postselection theorem \cite[Theorem 1]{Christandl_2009} gives an upper bound on this expression in terms of a purification of so-called de Finetti states.


This allows us to relate the $\epsilon$-secrecy of the QKD protocol for an arbitrary state $\rho_{ABE}$ to the $g_{n,x} \epsilon$-secrecy of the same protocol if the input is a purification of a probabilistic mixture of i.i.d. states. The final step of the postselection method then comprises relating this to the secrecy of an i.i.d. state. The (corrected) postselection Theorem (Corollary 3.1 in \cite{Nahar_2024}) states the following: Providing that one has proven security of a QKD protocol for i.i.d. states, yielding a key of length $\ell$
\begin{align*}
    &\frac{1}{2}\left|\left| \left(\left( \Phi^{(\ell)} - \Phi_{\mathrm{ideal}}^{(\ell)}\right) \otimes \mathrm{id}_E^n \right) \left(\tau_{ABE}^{\otimes n} \right) \right|\right|_1 \leq \epsilon_{\mathrm{sec}}
\end{align*}
with security parameter $\epsilon_{\mathrm{sec}} = \max\{\epsilon_{\mathrm{AT}},\epsilon_{\mathrm{PA}}\}$ (which is what we have just accomplished before), the security of the same protocol with a purification of a de Finetti state as input that hashes to a length $\ell' = \ell - 2 \log_2\left(g_{n,x}\right)$, where $g_{n,x}= {n+x-1 \choose n}$, is then
\begin{align}\label{eq:PS2}
    &\frac{1}{2}\left|\left| \left(\left( \Phi^{(\ell')} - \Phi_{\mathrm{ideal}}^{(\ell')}\right) \otimes \mathrm{id}_R \right) \left(\sigma_{A^nB^nR} \right) \right|\right|_1 \leq \epsilon_{\mathrm{sec}}'
\end{align}
secure, with security parameter $\epsilon_{\mathrm{sec}}' = \epsilon_{\mathrm{PA}} + 2 \sqrt{2 \epsilon_{\mathrm{AT}}}$. Note that the new postselection theorem by Ref.~\cite{Nahar_2024} straightforwardly extends from prepare-and-measure protocols to entanglement-based protocols.

Combining those findings yields the security statement against coherent attacks, which is summarized in the following theorem.

\begin{theorem}[\textbf{Security statement for coherent attacks}]\label{thm:SecurityStatementCoh}
    Let $\epsilon_{\mathrm{EV}}, \epsilon_{\mathrm{PA}}, \epsilon_{\mathrm{AT}}, \bar{\epsilon} >0$. Let $N$ be the total number of quantum signals exchanged, $k_T$ be the number of test rounds, and $n:=N-k_T$. The objective general high-dimensional QKD protocol is $\epsilon_{\mathrm{EV}} +  g_{n,x} \left( \epsilon_{\mathrm{PA}} + 2 \sqrt{2 \epsilon_{\mathrm{AT}}}\right)$-secure against coherent attacks, if, in case the protocol does not abort, the key is hashed to a length of
\begin{equation}
    \begin{aligned}
       \ell' \leq& n \min_{\tau_{ABE} \in \mathcal{S}^{\mathrm{AT}}} H_{\mathrm{min}}(X|E)_{\Phi(\tau_{ABE})}\\ 
      &~~~~~- \mathrm{leak}_{\mathrm{IR}} - 2 \log_2\left(\frac{1}{\epsilon_{\mathrm{PA}}}\right)- 2 \log_2\left(g_{n,x}\right),
    \end{aligned}
\end{equation}
where $g_{n,x} = {n+x-1 \choose n}$ and $\mathcal{S}^{\mathrm{AT}}$ is the acceptance set defined by the acceptance test in Theorem \ref{Thm:AcceptanceTest}.
\end{theorem}

In this work, we use $g_{n,x} \leq \left(\frac{e(n+x-1)}{x-1}\right)^{x-1}$ from Ref.~\cite{Nahar_2024} as an upper bound. In non-technical terms, Theorem \ref{thm:SecurityStatementCoh} allows us to relate the security against general attacks to the security against i.i.d. collective attacks, which we established earlier in Theorem \ref{thm:SecurityStatement}. While we still can relate the achievable secure key length $\ell'$ to a similar optimisation problem as we found for collective attacks, the stronger security claim comes at a price. In order to obtain secure key, the additional correction term $\log_2\left(g_{n,x}\right)$ tells us that we need to hash our raw key a bit more, while additionally, the total security parameter grows.

When we present our results, we plot secure key rates instead of achievable secure key lengths; thus, we divide both sides by $N$ and obtain

\begin{equation}
    \begin{aligned}
       \frac{\ell'}{N} \leq& \frac{n}{N} \min_{\tau_{ABE} \in \mathcal{S}^{\mathrm{AT}}} H_{\mathrm{min}}(X|E)_{\Phi(\tau_{ABE})} \\ 
      &~~~~~~- \delta_{\mathrm{leak}} - \frac{2}{N} \log_2\left(\frac{1}{\epsilon_{\mathrm{PA}}}\right)- \frac{2}{N} \log_2\left(g_{n,x}\right),
    \end{aligned}
\end{equation}
where $\delta_{\mathrm{leak}} := \frac{\mathrm{leak}_{\mathrm{IR}}}{N}$.


\section{Variable-Length Security}\label{sec:VarLengthSecurity}
In the previous section, we have proven the security of a high-dimensional QKD protocol that produces a key of fixed length or aborts. The most crucial part of such a fixed-length protocol in reality is to construct an acceptance set, which practically boils down to defining the expected channel behavior ($r_X$, potentially along with some margin $t_X$). This is because in many practical applications, the expected channel behaviour is not known or can change rapidly, such as in free-space and satellite-based QKD applications, which are two of the prototypical use-cases where high-dimensional QKD can play out its strengths. However, according to the security argument given in Section \ref{sec:SecurityArgument}, the accepted statistics have to be fixed before the run of the protocol, and the abort decision is made by comparing the actual measurement results with the acceptance set. For unknown or rapidly changing values, this would mean that even protocols with non-unique acceptance abort very often (and unique acceptance protocols basically all the time), leading to a low expected key rate \cite{Kanitschar_2023, Tupkary_2024}. The only possible counter-action is increasing the parameters $t_X$ to enlarge the acceptance set. This, however, diminishes the secure key length, as the reported length has to be compatible with the worst-case state within the given set, while not increasing the acceptance probability for varying channels significantly. 

We address this problem by introducing a variable-length version of our protocol and proving security, applying the theory by Ref. \cite{Tupkary_2024}. We start by stating the required protocol modifications and introducing some additional notation. 

The state generation, measurement, the announcement \& sifting phase as well as the key map of the protocol stated in Section \ref{sec:ProtocolDescription}, representing steps 1) - 3) and 5) remain untouched, while the main modifications pertain to steps 4), 6) and 7). The modified protocol steps read as follows.

\textbf{\textit{4'.) Variable-Length Decision.}} The communicating parties disclose a randomly chosen subset consisting of $k_T$ of their measurement results over the public channel and, depending on their observations, determine the required EC and hash length.

\textbf{\textit{6'.) Error-Correction \& Error-Verification}} On the remaining $N-k_T$ rounds, Alice and Bob perform error-correction to reconcile their raw keys $X$ and $Y$. The size of the error-correction leakage is already fixed by step 4'), where they decided to produce a key of length $\ell_i$ with EC leakage $\lambda_i^{\mathrm{EC}}$. As for the fixed-length protocol, they proceed with error-verification, where Alice chooses a two-universal hash function that hashes to $\left\lceil \log_2\left(\frac{1}{\epsilon_{\mathrm{EV}}}\right) \right\rceil$ bits and sends them to Bob, who performs the same hash function and compares them. In case they obtain the same hash values, they proceed.

\textbf{\textit{7'.) Privacy Amplification.}} Alice chooses a two-universal random hash function from $n$ to $\ell_i$ bits to to decouple Eve from their shared randomness.

For the fixed-length version of the protocol, the value of $\ell_i$ and $\lambda_i^{\mathrm{EC}}$ had to be fixed before the execution of the protocol; now this decision is made based upon the data. Let $\Vec{F}^{\mathrm{obs}} \in \mathbb{R}^{|\Theta|}$ denote the vector containing the experimental observations for all considered observables $X \in \Theta$ and denote $\Phi_X(\rho) := \Tr{\rho X}$.  

\subsection{Security against collective attacks}
\begin{figure}
\includegraphics[width=0.49\textwidth]{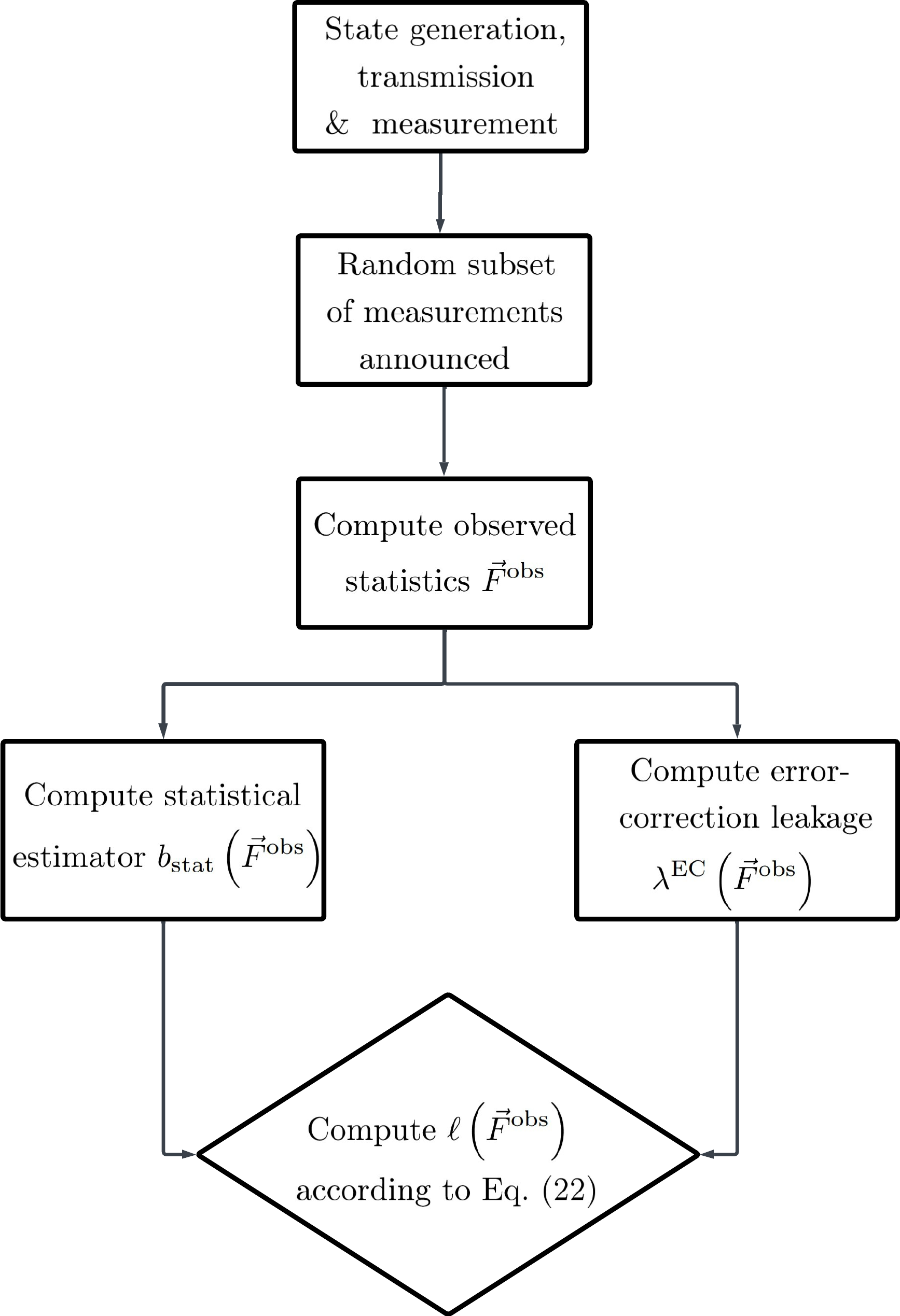}\\
\caption{Flowchart illustrating the variable-length decision. The length of the key is now a function of the observed statistics $\vec{F}^{\mathrm{obs}}$.\label{fig:flowchart}}
\end{figure}

In what follows, we summarize the idea of the variable-length security proof from Ref. ~\cite{Tupkary_2024} and adapt it to the objective high-dimensional QKD protocol. 
First, based on their observations $\Vec{F}^{\mathrm{obs}}$, Alice and Bob need to find a statistical estimator $b_{\mathrm{stat}}\left(\Vec{F}^{\mathrm{obs}}\right)$, which is a high-probability lower bound to the Rényi entropy $H_{\alpha}\left(Z^n|C^nE^n\right)_{\rho}$ of the underlying quantum state with, where $C$ denotes the register collecting the leaked classical information,
\begin{equation}
    \mathrm{Pr}\left[ b_{\mathrm{stat}}\left(\Vec{F}^{\mathrm{obs}}\right) \leq H_{\alpha}\left(Z^n|C^nE^n\right)_{\rho} \right] \geq 1-\epsilon_{\mathrm{AT}}.
\end{equation}
Intuitively, this estimator provides a high probability lower-bound on Eve's information about the key, based on the observed statistics $\vec{F}^{\mathrm{obs}}$. Finding such an estimator requires the construction of a set $V\left(\vec{F}^{\mathrm{obs}}\right)$ that contains the (unknown) quantum state responsible for Alice's and Bob's observations with high probability, i.e., $\mathrm{Pr}\left[ \tau_{AB} \in  V\left(\Vec{F}^{\mathrm{obs}}\right)\right] \geq 1-\epsilon_{\mathrm{AT}}$. Based on this estimator and the public announcements, Alice and Bob determine the information leakage $\lambda^{\mathrm{EC}}\left(\Vec{F}^{\mathrm{obs}}\right)$, which is an integer number between $0$ and $\lambda_{\mathrm{max}}^{\mathrm{EC}}$ quantifying how many bits of the key Eve learns by listening to Alice's and Bob's communication over the classical channel. An illustration of the variable-length decision can be found in Figure \ref{fig:flowchart}.

This allows calculating the secure key length as the difference between the statistical estimator and the error correction leakage, and some additional correction terms of lower order
\begin{equation}
\begin{aligned}
    &\ell\left(\Vec{F}^{\mathrm{obs}}\right) :=\\
    & \max\left\{0, \left\lfloor b_{\mathrm{stat}}\left(\Vec{F}^{\mathrm{obs}}\right) - \lambda^{\mathrm{EC}}\left(\Vec{F}^{\mathrm{obs}}\right) - \theta(\alpha, \epsilon_{\mathrm{PA}}, \epsilon_{\mathrm{EV}}) \right\rfloor \right\}
\end{aligned}    
\end{equation}

In the fixed-length scenario, the acceptance test was built upon defining expected statistics and finding the set of density matrices that could have produced these statistics only with some small probability; the complement of this set was the set $\mathcal{S}_{\mathrm{AT}}$. In the variable-length scenario, the construction of the acceptance set follows a slightly different logic. A fixed but otherwise unknown quantum state gives rise to observed statistics $\vec{F}^{\mathrm{obs}}$, which is a random variable. Based on this observation, we aim to construct a set $V(\vec{F}^{\mathrm{obs}})$ such that $\mathrm{Pr}\left[\tau_{AB} \in V(\vec{F}^{\mathrm{obs}})\right] \geq 1-\epsilon_{\mathrm{AT}}$. We are using observables, thus we have to adapt the argument from \cite{Tupkary_2024}. Let $\Theta$ denote the set of observables used in the protocol and let $x := ||X||_{\infty} = ||X||_{\infty}||\rho||_1 \geq ||X \rho ||_1$ for $X\in \Theta$. For each $X \in \Theta$, Hoeffding's inequality \cite{Hoeffding_1963} tells us 
\begin{equation}
   \mathrm{Pr}\left[ \left| F_X^{\mathrm{obs}} - \mathbb{E}[X] \right| \geq \mu_X \right] \leq 2 e^{-\frac{2 m_X \mu_X^2}{4 x^2}} =: \epsilon_X, 
\end{equation}
where $m_X$ is the number of rounds used for testing observable $X$. We apply Hoeffding's inequality to every observable $X \in \Theta$ and combine them, using Boole's inequality,
\begin{equation}
   \begin{aligned}
       \mathrm{Pr}\left[\bigcup_{X \in \Theta} \mathbb{E}\left[X\right] \notin \left[F_X^{\mathrm{obs}} - \mu_X, F_X^{\mathrm{obs}} + \mu_X\right] \right]\leq \sum_{X \in \Theta} \epsilon_X =: \epsilon_{\mathrm{AT}}. 
   \end{aligned} 
\end{equation}
In reverse, this means
\begin{equation}
    \mathrm{Pr}\left[\bigcap_{X \in \Theta} \mathbb{E}\left[X\right] \in \left[F_X^{\mathrm{obs}} - \mu_X, F_X^{\mathrm{obs}} + \mu_X\right] \right]\geq 1 - \epsilon_{\mathrm{AT}}
\end{equation}
and thus
\begin{equation}
    \mathrm{Pr}\left[ \rho \in V(\vec{F}^{\mathrm{obs}}) \right] \geq 1-\epsilon_{\mathrm{AT}},
\end{equation}
where $\vec{F}^{\mathrm{obs}}$ is defined as the vector containing $F_X^{\mathrm{obs}}$ in it's $X$-th component. This leads to
\begin{equation}
 \begin{aligned}
    &V\left(\vec{F}^{\mathrm{obs}}\right) :=  \left\{ \sigma \in \mathcal{D}(\mathcal{H}_A \otimes \mathcal{H}_B \otimes \mathcal{H}_E):\right. \\
    & \hspace{18mm} \forall X \in \Theta, |\Tr{\sigma X} - r_{X}| \leq \mu_{X}  \}.
\end{aligned}   
\end{equation}
Note that in the present case, despite the different argument, the main difference to the acceptance set for the fixed-length scenario is the decomposition of $\epsilon_{\mathrm{AT}}$ as a sum of the single-observable epsilons $\epsilon_X$. This, however, arises from using Hoeffding's inequality and can differ more significantly for different sampling arguments.

\begin{theorem}[\textbf{Variable-length security statement for collective attacks}]\label{thm:SecurityStatementVariable}
    Let $\epsilon_{\mathrm{EV}}, \epsilon_{\mathrm{PA}}, \epsilon_{\mathrm{AT}} > 0$. Let $N$ be the total number of quantum signals exchanged, $k_T$ the number of test rounds, and $n:=N-k_T$.  Let $\Theta$ be the set of observables considered for the objective protocol and $\vec{F}^{\mathrm{obs}} \in \mathbb{R}^{|\Theta|}$ be the vector containing the observations corresponding to $\Theta$. Based on $\vec{F}^{\mathrm{obs}}$, define the statistical estimator
    \begin{equation}
        \begin{aligned}
            b_{\mathrm{stat}}\left(\vec{F}^{\mathrm{obs}}\right):=& n \min_{\tau_{AB} \in V\left(\vec{F}^{\mathrm{obs}}\right)} H_{\mathrm{min}}\left(X|E\right)_{\Phi_{\mathrm{var}}\left(\tau_{ABE} \right)}\\
            &- n (\alpha-1) \log_2^2\left( \dim(X)+1 \right)
        \end{aligned}
    \end{equation}
    where $1 < \alpha < 1+ \frac{1}{\log_2\left( 2\dim(X)+1 \right)}$ and $V\left(\vec{F}^{\mathrm{obs}}\right)$ is such that it contains the state that produced $\vec{F}^{\mathrm{obs}}$ at least with probability $1-\epsilon_{\mathrm{AT}}$.

    Then, the variable length protocol, conditioned on obtaining $\vec{F}^{\mathrm{obs}}$ during acceptance testing and conditioned on passing the error-verification step, that hashes to length 
    \begin{equation}
        \begin{aligned}
            &\ell\left(\Vec{F}^{\mathrm{obs}}\right) :=\\
    & \max\left\{0, b_{\mathrm{stat}}\left(\Vec{F}^{\mathrm{obs}}\right) - \lambda^{\mathrm{EC}}\left(\Vec{F}^{\mathrm{obs}}\right) - \theta(\alpha, \epsilon_{\mathrm{PA}}, \epsilon_{\mathrm{EV}}) \right\}
        \end{aligned}
    \end{equation}
    where $\theta(\alpha,\epsilon_{\mathrm{PA}}, \epsilon_{\mathrm{EV}}) := \frac{\alpha}{\alpha-1} \left( \log_2\left( \frac{1}{4\epsilon_{\mathrm{PA}}} + \frac{2}{\alpha}\right) \right) + \left\lceil \log_2\left(\frac{1}{\epsilon_{\mathrm{EV}}} \right) \right\rceil$, using $\lambda^{\mathrm{EC}}\left(\vec{F}^{\mathrm{obs}}\right)$ bits for error-correction is $\epsilon_{\mathrm{EV}}$-correct and $\epsilon_{\mathrm{AT}} + \epsilon_{\mathrm{PA}}$-secure.
\end{theorem}

Intuitively, Theorem \ref{thm:SecurityStatementVariable} states that we can obtain a qualitatively similar security claim as we obtained in the fixed-length scenario (see Theorem \ref{thm:SecurityStatement}) even if we construct our statistical analysis around the observed rather than the expected measurement outcomes. However, as outlined before the statement, this requires a different security argument and leads to different correction terms.

\subsection{Security of variable-length HD-QKD protocols against coherent attacks}

Similarly to the fixed length protocol, we can define a protocol map $\Phi^{\ell_1,...,\ell_M}_{\mathrm{var}}$ that describes the action of the variable-length protocol hashing to keys of lengths $\ell_1, ..., \ell_M$, where $M$ is the number of possible different key lengths. Furthermore, let $\Phi^{\ell_1,...,\ell_M}_{\mathrm{var, ideal}}$ be its ideal counterpart. According to Theorem \ref{thm:SecurityStatementVariable}, the variable-length protocol was shown to be secure against collective attacks,
\begin{align}
    & \frac{1}{2} \left|\left| \left(\Phi_{\mathrm{var}}^{(\ell_1,...,\ell_M)} - \Phi_{\mathrm{var, ideal}}^{(\ell_1,...,\ell_M)} \right) \left(\rho_{A^nB^nE^n}\right)\right|\right|_1 \leq \epsilon_{\mathrm{PA}} + \epsilon_{\mathrm{AT}},
\end{align}
and $\epsilon_{\mathrm{EV}}$-correct. By a similar argument as in the previous section, exploiting the permutation invariance of the objective high-dimensional QKD protocol (see Ref.~\cite{Nahar_2024} for details), one can show that providing the protocol is secure against collective attacks with security parameter $\epsilon_{\mathrm{PA}} + \epsilon_{\mathrm{AT}}$, the protocol that hashes keys to length $\ell_i' := \ell_i - 2 \log_2(g_{n,x}) - 2\log_2\left(\frac{1}{\tilde{\epsilon}}\right)$ instead of length $\ell_i$ upon $\Omega_i$ with security parameter $g_{n,x} \left(\sqrt{8 \left(\epsilon_{\mathrm{PA}} + \epsilon_{\mathrm{AT}}\right)} + \frac{\tilde{\epsilon}}{2}\right)$,
\begin{align}
    & \frac{1}{2} \left|\left| \left(\Phi_{\mathrm{var}}^{(\ell_1',...,\ell_M')} - \Phi_{\mathrm{var, ideal}}^{(\ell_1',...,\ell_M')} \right) \left(\rho_{A^nB^nE^n}\right)\right|\right|_1 \\
    &~~~\leq g_{n,x} \left(\sqrt{8 \left(\epsilon_{\mathrm{PA}} + \epsilon_{\mathrm{AT}}\right)} + \frac{\tilde{\epsilon}}{2}\right),
\end{align}
where $\Phi_{\mathrm{var}}^{(\ell_1',...,\ell_M')}$ is identical to $\Phi_{\mathrm{var}}^{(\ell_1,...,\ell_M)}$, except that it hashes to a shorter key and $\epsilon_{\mathrm{EV}}$-secret. We summarize these findings in the following theorem.

\begin{theorem}[\textbf{Variable-length security statement for coherent attacks}]\label{thm:SecurityStatementVariableCoh}
    Let $\epsilon_{\mathrm{EV}}, \epsilon_{\mathrm{PA}}, \epsilon_{\mathrm{AT}}, \tilde{\epsilon}> 0$. Let $N$ be the total number of quantum signals exchanged, $k_T$ the number of test rounds, and $n:=N-k_T$. Providing the objective high-dimensional variable-length QKD protocol is secure against collective attacks according to Theorem \ref{thm:SecurityStatementVariable} with security parameter $\epsilon_{\mathrm{PE}} + \epsilon_{\mathrm{AT}}$ and correctness parameter $\epsilon_{\mathrm{EV}}$, conditioned on obtaining $\vec{F}^{\mathrm{obs}}$ during acceptance testing and passing the error-verification, the protocol is secure against coherent attacks with security parameter $g_{n,x} \left( \sqrt{8 (\epsilon_{\mathrm{PE}} + \epsilon_{\mathrm{AT}})} + \frac{\tilde{\epsilon}}{2} \right)$, if the key is hashed to a length of
    \begin{equation}\label{eq:VarLengthCohKRFormula}
\begin{aligned}
            &\ell\left(\Vec{F}^{\mathrm{obs}}\right) :=\\
    & \max\left\{0, b_{\mathrm{stat}}\left(\Vec{F}^{\mathrm{obs}}\right) - \lambda^{\mathrm{EC}}\left(\Vec{F}^{\mathrm{obs}}\right) - \theta(\alpha, \epsilon_{\mathrm{PA}}, \epsilon_{\mathrm{EV}}) \right.\\
    &~~~~~~~~~~~~ \left.- 2 \log_2\left(g_{n,x}\right) - 2\log_2\left( \frac{1}{\tilde{\epsilon}}\right)\right\},
        \end{aligned}
    \end{equation}
where $\theta(\alpha,\epsilon_{\mathrm{PA}}, \epsilon_{\mathrm{EV}}) := \frac{\alpha}{\alpha-1} \left( \log_2\left( \frac{1}{4\epsilon_{\mathrm{PA}}} + \frac{2}{\alpha}\right) \right) + \left\lceil \log_2\left(\frac{1}{\epsilon_{\mathrm{EV}}} \right) \right\rceil$ and $g_{n,x} = {n+x-1 \choose n}$ for $x = d_A^2 d_B^2$, using $\lambda^{\mathrm{EC}}\left(\vec{F}^{\mathrm{obs}}\right)$ bits for error-correction.
\end{theorem}
Similar to the variable-length statement for collective attacks, Theorem \ref{thm:SecurityStatementVariableCoh} presents a variable-length version for the achievable secure key length against coherent attacks, which builds the statistical analysis around the observed rather than the expected behaviour of the channel. The core message is that we obtain a qualitatively similar security statement to the fixed-length case, even if we build our security argument upon the observed rather than the expected statistics.

\section{Security proof method} \label{sec:ProofMethod}
Having discussed the security for two different attack scenarios and both the fixed and variable-length scenarios, it remains to find a way to evaluate the given expressions. However, as discussed earlier and found in Ref.~\cite{BKPH_2023}, widely available convex optimization methods \cite{Araujo_2022, Winick_2018} cannot handle high-dimensional QKD protocols. On the other hand, computationally less demanding methods \cite{Sheridan_2010, Doda_2021} rely on hard-to-implement measurements, which for most practical setups are not available. Instead, we employ a recent technique developed in Ref.~\cite{Kanitschar_2024a} that provides lower bounds on the secure key rate without high computational demands, solely relying on available measurements. 
\subsection{Key rate evaluation}
While this method computes asymptotic key rates, the formulation of our composable finite-size key rate finding problems derived in Sections \ref{sec:SecurityArgument} and \ref{sec:VarLengthSecurity} in terms of the min-entropy already allows us to directly apply this framework. The basic idea of this method is the following \cite{Kanitschar_2023}. It is well known that the min-entropy can be expressed as the negative binary logarithm of Eve's probability of guessing Alice's key string correctly, $H_{\mathrm{min}}(X|E)_{\rho} = - \log_2\left(p_{\mathrm{guess}}\right)$. This allows us to recast the minimization of the min-entropy over $\mathcal{S}^{\mathrm{AT}}$ as maximization over all possible measurements Eve might carry out in order to guess Alice's outcomes. Applied to our finite-size optimization problem, this turns into
\begin{equation}\label{eq:Primal}
   \begin{aligned}
    p_{\mathrm{guess}}& = \max_{\stackrel{\rho_{ABE}}{\{E^{l}\}_{l=0}^{d-1}}} \sum_{l,k} \Tr{\rho_{ABE} A_1^{l} \otimes B_1^{k} \otimes E^{l}}\\
    \mathrm{s.t.: }& \\
    & \sum_{l} E^{l} = \mathbbm{1},& \\
    &  \Tr{\rho_{ABE}} = 1,&\\
    & \langle \hat{W}_0 \rangle - \mu_0 \leq \Tr{\left(\hat{W}_0\otimes \mathbbm{1}_E\right) \rho_{ABE}} \leq \langle \hat{W}_0 \rangle + \mu_0,\\
    & \Tr{\left(\hat{W}^{(i)}_j\otimes \mathbbm{1}_E\right) \rho_{ABE}} \leq \langle \hat{W}_j\rangle + \mu_j,\\
    & E^{l} \geq 0,\\
    &\rho_{ABE} \geq 0,
\end{aligned} 
\end{equation}
for $j\in \{1,..., N_{W}\}$ and $l \in \{ 0,..., d-1\}$. The observables $\hat{W}_j$ are detailed in Section \ref{sec:Results}. The numerical values of $\mu_j$ are calculated according to Theorem \ref{Thm:AcceptanceTest}. Instead of solving this semi-definite program directly, the method derives its dual 
\begin{equation}\label{eq:Dual}
\begin{aligned}
 &\min_{y_0, z_0^L, z_0^U, \{z_j\}_{j=1}^{N_{W}}} y_0 + z_0^U \left( \langle\hat{W}_0 \rangle + \mu_0 \right) - z_0^L \left( \langle \hat{W} \rangle - \mu_0\right)\\
 &~~~~~~~~~~~~~~~~~~~~~+ \sum_{j=1}^{N_{W}} z_j \left( \langle \hat{W}_j\rangle + \mu_j\right)  \\
   &~~~~\mathrm{s.t. }\\
   &~~~~~~~ y_0 \geq \lambda_{\mathrm{max}}\left(M_{\ell}\right) ~~~ \forall \ell=0,...,d-1 \\
   &~~~~~~~ z_0^L, z_0^U, z_j\geq 0 ~~~ \forall j=1,...,N_{W}\\
   &~~~~~~~y_0\in \mathbb{R},
   \end{aligned}
\end{equation}
which boils down the problem to solve a rather simple minimization problem subject to a constraint that depends on the maximal eigenvalue of a matrix which is a function of the chosen observables $M_{\ell}:=\ketbra{\ell} \otimes \mathbbm{1}_d - \left(z_0^U - z_0^L \right) \bar{W}_0 - \sum_{j=1}^{N_{W}}z_j^U\bar{W}_j$. 

Thus, the crucial task was reduced to finding experimentally accessible observables. This can be significantly simplified by extending the set of accessible operators using a matrix-completion technique \cite{Tiranov_2017, Martin_2017}. For further details, and a discussion about the complexity scaling, we refer the reader to Ref. \cite{Kanitschar_2024a}.

\subsection{Formalization of the protocol}\label{sec:PPMap} 
For illustration purposes, we apply our method to a high-dimensional temporal entanglement setup using a Franson interferometer, analyzed in Ref. \cite[Protocol 1]{BKPH_2023}. A source prepares polarized photon pairs of the form $\ket{\Psi_1} = \ket{\mathrm{DD}}\otimes \frac{1}{\sqrt{d}}\sum_{k=0}^{d-1}\ket{kk}$ which are distributed to the communicating parties. With probability $p_1$ each party performs a Time-of-Arrival (ToA) measurement, represented by $\{A_1^x\}_{x=0}^{d-1}$ and $\{B_1^y\}_{y=0}^{d-1}$, respectively, and with probability $1-p_1$ each of them performs a temporal superposition (TSUP) measurement of not necessarily neighboring time-bins, represented by $\{A_2^x\}_{x=0}^{d-1}$ and $\{B_2^y\}_{y=0}^{d-1}$, respectively. We start by deriving the protocol map representing the action of this protocol, specifying the map $\Phi$ introduced in Section \ref{sec:SecurityArgument}, following the formalism in \cite{Lin_2019, Upadhyaya_2021}. 

For each round, Alice and Bob may store a measurement outcome $\alpha_i$ and $\beta_i$ and make a public announcement $a_i$ and $b_i$, where the measurement outcomes are drawn from the sets $\mathcal{X}_A$ and $\mathcal{X}_B$ and the announcements from the set $\mathcal{A}_A$ and $\mathcal{A}_B$ for Alice and Bob, respectively. In our case, these announcements are the choice of measurement, i.e., a number $b$ related to the label of the POVM measurement performed. For the analyzed protocol, $b=1$ corresponds to time-of-arrival measurements, labeled `ToA', while $b=2$ refers to superposition measurements of neighboring time bins, labeled `TSUP'. They store their measurement outcomes in their private registers $\bar{A}$ and $\bar{B}$ and their announcements in their public registers $\tilde{A}$ and $\tilde{B}$. Since we perform direct reconciliation, the key map $g$ saves the outcome of the map $g:~\mathcal{X}_A \times \mathcal{A}_A \times \mathcal{A}_B \rightarrow \{0, ..., d-1, \perp\}$ to the key register $Z$. In more detail, since the protocol only generates a key when both Alice and Bob measure ToA, the key map for our protocol is trivial and simply copies Alice's measurement outcome contained in her private register $\bar{A}$ to the key register $Z$. This sets the stage for introducing the formal protocol map. Let $\rho_{AB}$ be Alice's and Bob's shared quantum state after the generation and distribution of the high-dimensional entangled state, and let $\rho_{ABE}$ denote Eve's purification. Furthermore, since Alice's and Bob's announcements are public, Eve has access to them. We use the short notation $[E]$ to label Eve's register, including all available public information. Eve's goal is to perform some (to us unknown) POVM measurement, aiming to gain as much information as possible about Alice's measurement outcome. Since Alice and Bob only generate a key if both measure the time of arrival, we obtain the following map, representing their measurement and announcement steps

\begin{equation}
  \begin{aligned}
    &\Phi_{\mathrm{M\&A}}\left(\rho_{ABE}\right)\\
    &= \sum_{i,j} \ketbra{\mathrm{ToA}}_{\tilde{A}} \otimes \ketbra{i}_{\bar{A}} \otimes \ketbra{\mathrm{ToA}}_{\tilde{B}} \otimes \ketbra{j}_{\bar{B}} \\
    &\otimes \left[ \left( \sqrt{A_1^i} \otimes \sqrt{B_1^i} \otimes \sqrt{E^i} \right) \rho_{ABE} \left( \sqrt{A_1^i} \otimes \sqrt{B_1^i} \otimes \sqrt{E^i} \right)^{\dagger}\right],
\end{aligned}  
\end{equation}
and the key map isometry reads
\begin{align}
    V = \sum_k \ket{k}_Z &\otimes \ketbra{\mathrm{ToA}}_{\tilde{A}} \otimes \ketbra{\mathrm{ToA}}_{\tilde{B}} \otimes \ketbra{k}_{\bar{A}}.
\end{align}
Thus, the final state $\sigma_{Z,[E]}$ is obtained by
\begin{equation}
\begin{aligned}
    \sigma_{Z[E]} &= \mathrm{Tr}_{A\bar{A}B\bar{B}}\left[ V \Phi_{\mathrm{M\&A}} V^{\dagger} \right]. 
\end{aligned}  
\end{equation}

Nonetheless, since the announcements for this protocol are trivial, they do not give any additional information. Hence, after inserting the expressions for Alice's and Bob's key measurement POVM and tracing out the announcement registers, we obtain
\begin{widetext}
 \begin{align}
    \sigma_{ZE} &= \sum_{i,j} \ketbra{i}_Z \otimes \mathrm{Tr}_{AB}\left[ \left( \sqrt{\ketbra{i}} \otimes \sqrt{ \ketbra{j}} \otimes \sqrt{E_i} \right) \rho_{ABE} \left( \sqrt{\ketbra{i}} \otimes \sqrt{\ketbra{j}} \otimes \sqrt{E_i} \right)^{\dagger}\right] \\
    &= \sum_{i,j} \ketbra{i}_Z \mathrm{Tr}_{AB}\left[ \left(\ketbra{i}_A \otimes \ketbra{j}_B \otimes E^i\right) \rho_{ABE}  \right]\\
    &=: \Phi(\rho_{ABE}).
\end{align}   
\end{widetext}

The last line defines the final protocol map $\Phi$. For this $\Phi$ the objective function of the optimisation in Eq. (\ref{eq:Primal}) we use to calculate $H_{\mathrm{min}}(X|E)_{\Phi(\rho)} = -\log_2(p_{\mathrm{guess}})$ becomes\\
\begin{equation}
\begin{aligned}
    &p_{\mathrm{guess}} \\
    &= \max_{\stackrel{\rho_{ABE}}{ \{E^{l}\}_{l=0}^{d-1}}} \mathrm{Tr}_{ZE}\left[\sum_{i,j=0}^{d-1} \mathrm{Tr}_{AB}\left[ \left(\ketbra{i}_A \otimes \ketbra{j}_B \otimes E^i\right) \rho_{ABE}  \right]\right]\\
    &= \max_{\stackrel{\rho_{ABE}}{ \{E^{l}\}_{l=0}^{d-1}}} \sum_{i=0}^{d-1} \mathrm{Tr}\left[ \left(\ketbra{i}_A \otimes \mathbbm{1}_B \otimes E^i\right) \rho_{ABE}  \right].
\end{aligned}
\end{equation}
which then translates to the dual given in Eq. (\ref{eq:Dual}).

\section{Results}\label{sec:Results}
In this section, we demonstrate our method by deriving composable finite-size key rates for the high-dimensional temporal entanglement protocol and setup analyzed in Ref. \cite{BKPH_2023}. While this is one illustrative example, our method applies to general high-dimensional protocols in a very similar fashion. The underlying data can be found in Ref. \cite{Data}.

In accordance with Ref. \cite{Kanitschar_2024a}, we chose $d-1$ observables 
$\hat{W}_k:=q_0 \sum_{i=0}^{d-1} \ket{i,i}\!\!\bra{i,i} + \sum_{z=1}^{d-1}\sum_{i=0}^{d-1} q_{z,i} \left(\ket{i,i}\!\!\bra{i+z, i+z} + \ket{i+z,i+z}\!\!\bra{i, i}\right)$ and $\hat{W}_0:= \sum_{\stackrel{i,j=0}{i\neq j}}^{d-1} p \ket{i,j}\!\!\bra{i,j}$, where $p = 1$. We chose the $q_{z,i}$ such that the $k$-th operator acts only on the $k$-th off-diagonal, which allows us to include the optimal choice of the relative weights between the witnesses in the optimization problem. More details regarding this choice and the intuition behind it can be found in \cite[Appendix D]{Kanitschar_2024a}.

First, we cover the fixed-length scenario, which best suits theoretical discussions of achievable `one-shot' key rates, followed by a discussion of variable-length key rates under specific noise models. In all examples given, we use $r_S$ to refer to the splitting ratio of the beam splitters in Alice's and Bob's labs that passively choose the measurement setting (ToA or TSUP). For example, a splitting ratio of $r_S = 1\%$ means that $1\%$ of the photons are directed to the TSUP measurement apparatus, while the remaining $99\%$ are measured in the ToA setup. Thus, they obtain $r_S^2N$ rounds of aligned TSUP measurements and $(1-r_S)^2$ rounds of aligned ToA measurements. Given that $8(d-1)$ different click events are combined into $W_1$, we obtain $m_{W_1} := \frac{r_S^2 N}{8(d-1)}$. Additionally, we choose $m_{W_2} = m_{W_1}$ such that $n:= N (1-r_S)^2 - Nr_s^2 = N(1-2r_s)$, since the click elements required for $W_2$ are obtained from ToA measurements. 

For demonstration purposes, if not mentioned otherwise, we assume an isotropic noise model $\rho = v \ketbra{\Psi_{1}} + \frac{(1-v)}{d^2} \mathbbm{1}_{d^2}$. However, we want to emphasize that our method applies independently of the chosen noise model, and the noise model is solely required to simulate the observations for key rate plots. The security analysis is general and does not require any assumptions about the state or the noise model.

Also, note that the leakage in the information reconciliation phase $\mathrm{leak}_{\mathrm{IR}}$ appearing in the key rate formulas given in Section \ref{sec:SecurityArgument}, is the sum of the error-correction leakage $\mathrm{leak}_{\mathrm{EC}}$ and the length of the hash used for error-verification $\log_2\left( \frac{2}{\epsilon_{\mathrm{EV}}} \right)$. If not stated otherwise, we assume ideal error correction.

\begin{figure}
\subfloat[i.i.d. collective attacks. \label{fig:examination_N_collective} ]{
    \includegraphics[width=0.45\textwidth]{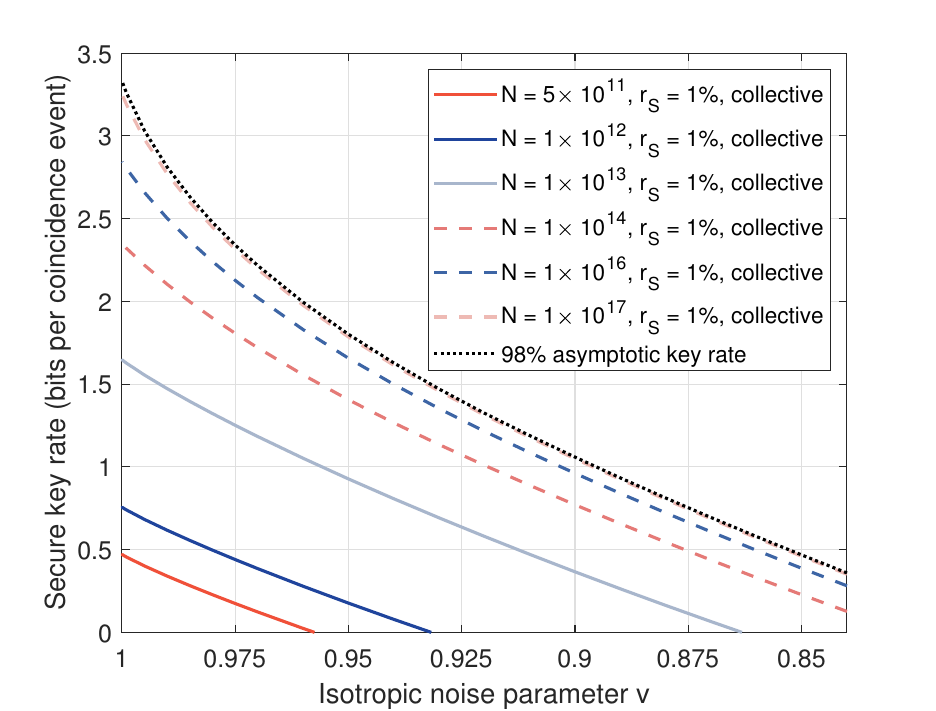}}\\
\subfloat[coherent attacks. \label{fig:examination_N_coherent}]{
    \includegraphics[width=0.45\textwidth]{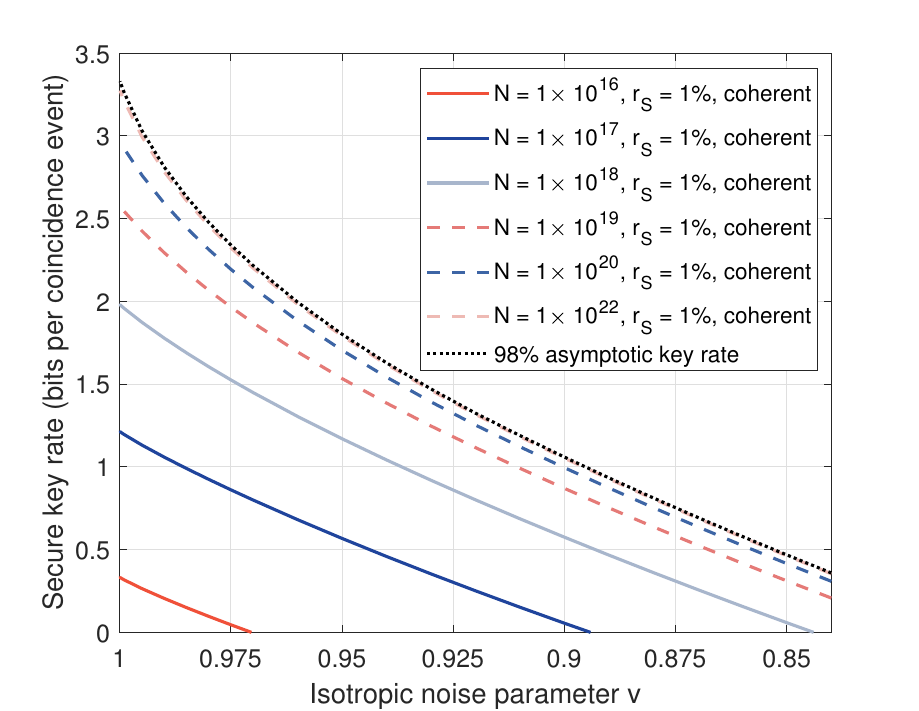}}
\makeatletter\long\def\@ifdim#1#2#3{#2}\makeatother
\caption{Composable secure key rates against (a) i.i.d. collective attacks and (b) coherent attacks for different block sizes $N$ and constant splitting ratio of $1\%$. The system dimension was fixed to $d=16$ and we used the witness $\mathrm{KH1}$.}
\end{figure}

\subsection{Fixed-length key rates}

We start with fixed-length keys, also called `one-shot' key rates, corresponding to key rates usually reported in the literature.  Recall that for the definition of the acceptance set (see Theorem \ref{Thm:AcceptanceTest}), one may choose to extend the set of accepted statistics by choosing the parameters $t_X > 0$ (for each observable $X$). Intuitively, this extends the set of accepted statistics from a point to a (hyper-) rectangle with sides $t_X$ for $X \in \Theta$. To simplify notation and to describe the choices for the acceptance test with a single parameter, we measure the chosen $t_X$ for the different observables in multiples of their respective $\mu_X$ by introducing a factor $t_F$, $t_X = t_F \mu_X$. In the whole subsection discussing the results for the fixed-length scenario, we fix $t_F = 1$. The choice of $t_F$ influences the one-shot key rate (larger $t_F$ extends the set of feasible states, thus lowering the key rate) and the acceptance probability of the acceptance test (the larger $t_F$, the higher the acceptance probability). We discuss the influence of $t_F$ in more detail in Section \ref{sec:VarLenKRresults}, when we compare fixed-length to variable-length key rates.

Since the choices for the epsilons can have an influence on the secure key rates, we discuss our considerations in more detail in Appendix \ref{APDX:ImpDetails}, where we express the security epsilons as functions of two real, positive parameters $r,s >0$. While we did not optimize our choices to the full extent, as this would go beyond the scope of this paper and is best discussed for real experiments, we categorize the different epsilons by their impact on the key rate and choose them accordingly. For the whole section, we set $r= \frac{1}{1000} = s$.

For the collective attack scenario, this leads to $\epsilon_{\mathrm{AT}} = 9.99\times 10^{-11}$, $\epsilon_{\mathrm{PA}} = 9.99\times 10^{-11}$ and $\epsilon_{\mathrm{EV}} = 1 \times 10^{-13}$, resulting in a total security parameter of $\epsilon = 1 \times 10^{-10}$. In Figure \ref{fig:examination_N_collective}, we examine the key rates for varying $N$ and compare the obtained finite-size key rates with the asymptotic rates for $N \rightarrow \infty$. We fixed the protocol dimension to $d=16$ and the splitting ratio to $r_S = 1\%$. This means for passive choice settings $r_S^2$ of the rounds are available for TSUP measurements on both sides, and $(1-r_S)^2 = 0.9801$ of the rounds are measured in ToA by both parties. Given our choice of $m_{W_1} = m_{W_2}$, $(1-2r_S) = 98\%$ of all rounds are available for key generation. Note that the comparable asymptotic key rates refer to $N\rightarrow \infty$, where the testing fraction can be chosen to be infinitesimally small. Thus, to ensure a fair comparison, we need to adjust the asymptotic rates by a factor of $0.98$ to take the testing into account. We plot secure key rates for $N \in \{5\times 10^{11}, 1\times 10^{12}, 1\times 10^{13}, 1\times 10^{14}, 1\times 10^{16}, 1\times 10^{17}\}$ and observe convergence to the adjusted asymptotic key rate curve, as expected. We obtain nonzero finite-size key rates for block sizes of $N=5 \times 10^{11}$ and above. For $N=10^{17}$, our finite-size results effectively reassemble the asymptotic key rates. We note that the comparably high number of rounds required to obtain a nonzero key is due to the low testing ratio, which we chose to examine asymptotic behaviour (as asymptotic rates only require infinitesimal testing). We will examine higher testing ratios for realistic block sizes later.

For coherent attacks, the security parameters scale according to Theorem \ref{thm:SecurityStatementCoh}. Again, we discuss our parameter choices in detail in Appendix \ref{APDX:ImpDetails}, and chose $r=\frac{1}{1000}=s$.

We aim to keep $\nu_{\mathrm{AT}}$ as large as possible at the cost of lowering the other security parameters more. We choose $\nu_{\mathrm{AT}} = 9.98 \times 10^{-11}$, $\nu_{\mathrm{PA}} = 9.99\times 10^{-14}$, and $\nu_{\mathrm{EV}} = 1\times 10^{-13}$ such that they add up to a total adjusted security parameter of $\nu = 10^{-10}$. The corresponding epsilon parameters can be obtained by scaling according to the security statement in Theorem \ref{thm:SecurityStatementCoh}.

In Figure \ref{fig:examination_N_coherent}, we plot the obtained secure key rates vs. visibility and examine the asymptotic behavior for large $N$ of the secure key rates against coherent attacks. We fix the dimension again to $d=16$ and the splitting ratio $r_S = 1\%$ and plot the curves for $N \in \{1\times 10^{16}, 1\times 10^{17}, 1\times 10^{18}, 1\times 10^{19}, 1\times 10^{20}, 1 \times 10^{22}\}$. We obtain nonzero key rates for $N=10^{17}$ and above and observe convergence to the adjusted asymptotic rate $N\rightarrow \infty$ for large $N$. Again, we note that we chose the low splitting ratio to examine the asymptotic behavior. Since the asymptotic rate does not take testing into account, we again rescale by the corresponding factor of $0.98$. We obtain nonzero key rates for significantly lower $N$ for higher splitting ratios.

\begin{figure}
\includegraphics[width=0.45\textwidth]{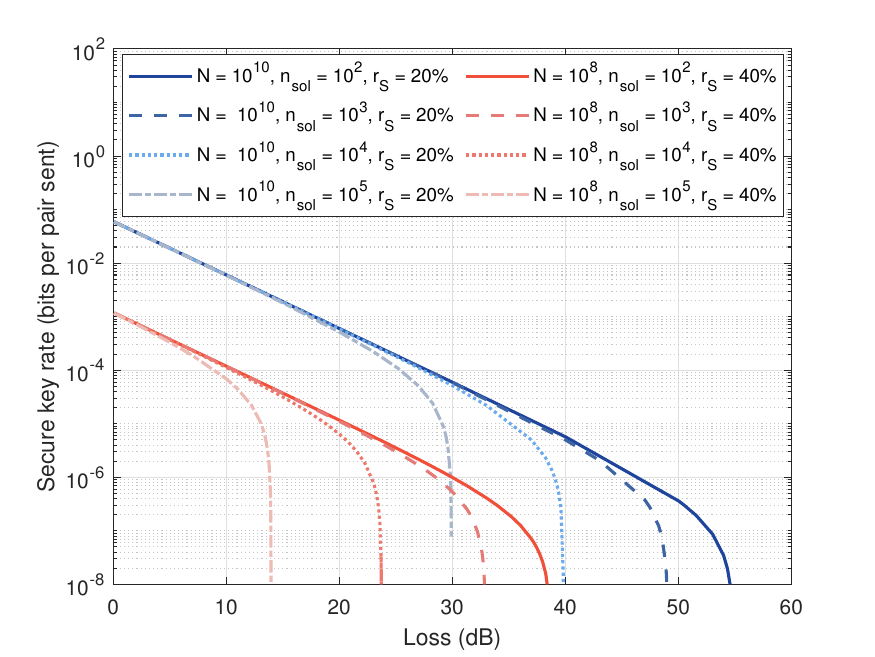}\\
\caption{Composable secure key rates against i.i.d. collective attacks versus loss for different block sizes $N$, splitting ratios $r_s$ and noise levels, parametrized by the number of solar photons $n_{\mathrm{sol}}$. The system dimension was fixed to $d=16$. \label{fig:KR_vs_Loss_coll}}
\end{figure}

Next, we examine the behaviour with respect to loss for different block sizes $N$, splitting ratios $r_s$, and different noise levels. We use the noise model from Ref. \cite{BKPH_2023} and choose the production rate $c_{\mathrm{prod}} = 0.1$ per time frame for a time frame length of $T=5.4\times 10^{-9}$. The dark-count rate was assumed to be $100$ per second, and the detection efficiency $\eta_D = 0.90$. In Figure \ref{fig:KR_vs_Loss_coll}, we plot the obtained secure key rates (in bits per pair sent) versus loss. Curves in different shades of blue correspond to $N=10^{10}$ with a splitting ratio to $r_s = 20\%$, and curves in different shades of red correspond to $N=10^{8}$ with a splitting ratio to $r_s = 40\%$. Different line styles (solid, dashed, dotted, and dot-dashed) indicate different solar photon rates $n_{\mathrm{sol}} \in \{10^2, 10^3, 10^4, 10^5 \}$. We observe that for $N=10^{10}$ and $n_{\mathrm{sol}} = 10^2$, corresponding to almost absolute darkness, we achieve a nonzero key rate up to about $51$dB channel attenuation, which is just shy of the asymptotic rates. For $N=10^8$, the tolerable loss decreases to $38.5$dB. While the achievable loss decreases with increasing solar photon rate, in both cases, we still obtain a nonzero key for $n_{\mathrm{sol}} = 10^5$, corresponding to sunrise \cite{Bulla_2022}. Increasing the splitting ratio $r_S$ can improve the noise tolerance even further.
This demonstrates that high-dimensional QKD can play a crucial role in overcoming noise in loss regimes present in satellite-based QKD. 

\begin{figure}
\includegraphics[width=0.45\textwidth]{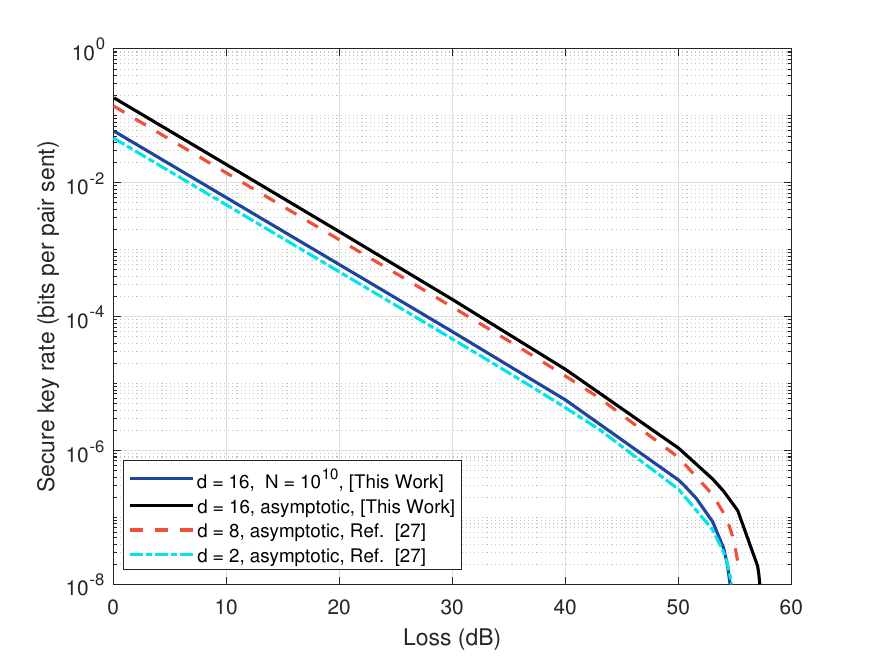}\\
\caption{Comparison of our results with asymptotic behaviour and numerical key rate curves according to Ref. \cite{BKPH_2023}. The splitting ratio was fixed to $r_s = 20\%$ and the noise level, parametrized by the number of solar photons, was chosen to be $n_{\mathrm{sol}} = 10^2$. The system dimension varies and is given in the legend. \label{fig:KR_vs_Loss_comp}}
\end{figure}

As shown in Figure \ref{fig:KR_vs_Loss_comp}, already for $N=10^{10}$, the collective attack key rates (solid blue curve) are only slightly below the asymptotic rates (solid black curve) in terms of tolerable loss and less than half an order of mangitude lower in terms of achievable secure key rate. Additionally, we compare our results with the key rates from Ref. \cite{BKPH_2023} obtained for the same setup, using the same noise model, but relying on a convex optimisation Gauss-Radau method \cite{Araujo_2022}. While Ref. \cite{BKPH_2023} only reports asymptotic rates, similar finite-size corrections are expected, so the comparison of asymptotic curves allows similar conclusions about finite-size results.
In Ref. \cite{BKPH_2023}, the maximum dimension feasible on a standard computer was $d=8$ (and $d=12$ using a cluster), with the $d=8$ curve (dashed orange) requiring several days of computation. By contrast, our method produces the $d=16$ curve (solid black) within minutes. Already at $d=16$ our method outperforms the numerical $d=8$ result in both key rate and tolerable loss, thereby overcoming the apparent drawback of using $H_{\mathrm{min}}$ instead of the von Neumann entropy. This highlights the clear advantage of our method, which unlocks the use of genuine high dimensions without substantial compromises on the secure key rate.

\subsection{Variable-length key rates}\label{sec:VarLenKRresults}
Next, we come to the variable-length scenario. To illustrate the advantage of finite-size key rates, we need to employ a noise model that translates certain physical parameters of the devices used, such as the pair-production rate of the source, the dark-count rate, and efficiency of the detector and channel parameters like loss and the number of environmental photons (for example solar photons) into coincidence click matrices. The latter can be used to derive the observations for our witness operators (observables). For illustration purposes, we use the noise model developed in Ref. \cite{BKPH_2023}, although, of course, the arguments apply to any noise model. Additionally, one needs to simulate natural channel fluctuations, as they appear under any realistic conditions. In Appendix \ref{APDX:SimChannel}, we discuss and model two different types of channels that may appear in real experiments. We call the first one `Statistically Fluctuating Channel', which refers to a well-characterized, stable channel with channel parameters probability of photon loss $p_L$ and a number of excess (solar) photons $n_{\mathrm{sol}}$ that follow a normal distribution around certain characterized means. This represents a stable atmospheric channel under constant weather conditions. The second channel is called the 'Rapidly Fluctuating Channel` and describes an atmospheric channel under unstable weather conditions, such as cloudy or foggy nights, or changing brightness, e.g., due to the rising sun. We apply a simplified model and assume that randomly one out of nine Statistically Fluctuating Channels with different means $\mu_{\mathrm{sol}}$ and $\mu_{\mathrm{loss}}$ manifests. This, of course, is only for demonstration purposes, while in reality, a plethora of conditions may apply. While the second scenario is more realistic, the first one already suffices to demonstrate the clear advantage of variable-length protocols. Throughout the whole chapter, we fix the dimension to $d=16$. We direct the interested reader to Appendix \ref{APDX:ImpDetails} for details about our choices for the security parameters. In this section, we set $r=\frac{1}{1000}=s$, which turns out to be a sufficiently good choice.

\begin{figure}
\subfloat[Comparing variable-length and fixed-length key rates for $100$ simulated runs. \label{fig:varLengthKRs_vs_Runs} ]{
    \includegraphics[width=0.45\textwidth]{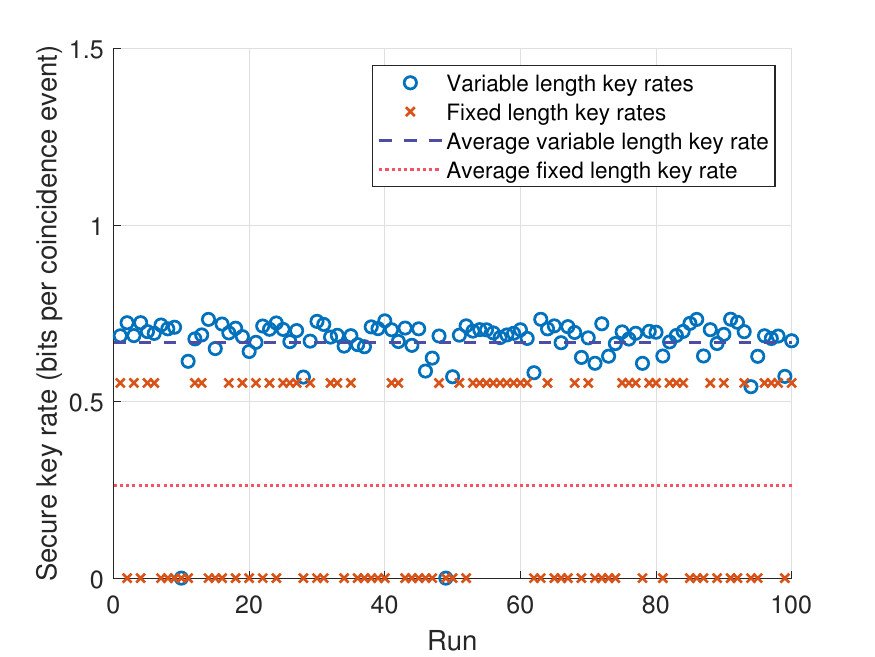}}\\
\subfloat[Key rate vs. $t_F$. \label{fig:varLengthKRs_vs_t}]{
    \includegraphics[width=0.45\textwidth]{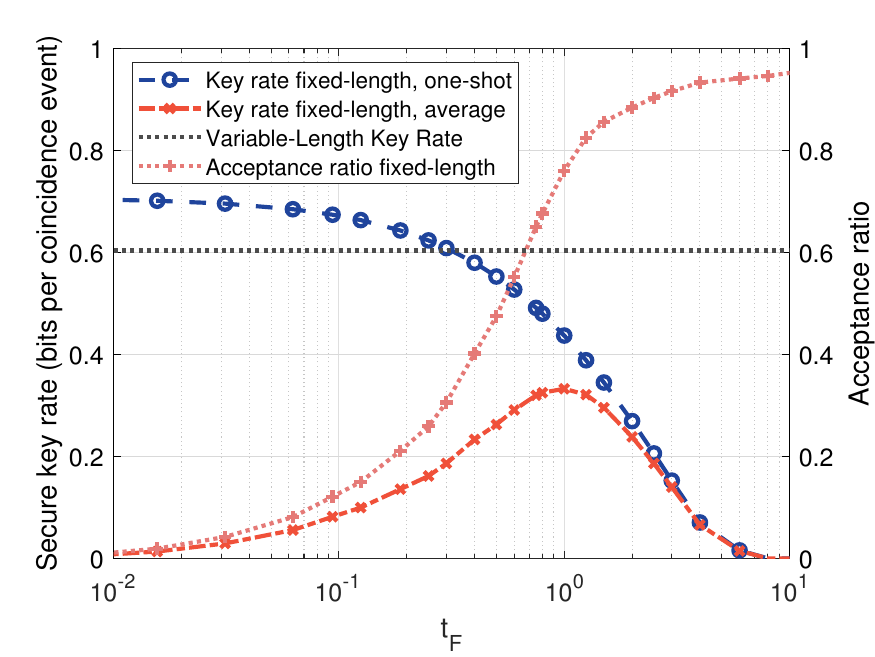}}
\makeatletter\long\def\@ifdim#1#2#3{#2}\makeatother
\caption{Variable-length key rates for the `Statistically Fluctuating Channel'. The system dimension was fixed to $d=16$ }
\end{figure}

We start our discussion by reviewing the notion of acceptance testing for fixed-length protocols. There, before the protocol is executed, one has to determine and fix an expected channel behaviour by determining expected observations for the witness operators. This is usually a single data point. In this case, of course, every tiny deviation from the expected behaviour leads to a deviation from the observed statistics. During the acceptance test, one compares these observed statistics with the expected statistics, and it aborts if they do not match. Therefore, under this `unique-acceptance' scenario, one would abort the protocol almost all the time, making it secure but essentially useless. We can fix this problem by introducing some margin to extend the set of accepted statistics (see parameter $t_X$ in Theorem \ref{Thm:AcceptanceTest}). As elaborated on earlier, by doing so, we build a multi-dimensional interval centered on our original expected statistics, which increases the probability of acceptance. This, however, comes at the cost of extending the set of states that pass the acceptance set, hence the set of states that potentially produced the observed statistics, and thus lowers the key rate. To ease the illustration, we parametrize $t_X$ for each observable $X \in \Theta$ by a positive real number $t_F$, $t_X = t_F \mu_X$. In contrast, the variable-length security argument is built around the observed statistics and \textit{not} around the expected statistics and adapts the secure key length to the quality of the observations. Therefore, for variable-length key rates, we can set $t_F=0$. As derived in Section \ref{sec:VarLengthSecurity}, the acceptance testing security parameter $\epsilon_{\mathrm{AT}}$ differs between the fixed- and the variable-length scenarios, where $\epsilon_{\mathrm{AT}}$ is the sum of the acceptance-testing epsilons $\epsilon_X$ for each single observable $X\in\Theta$. To simplify application for the variable-length scenario, from now on, we require that all the $\epsilon_X$ are equal, hence $\epsilon_{\mathrm{AT}} = |\Theta| \epsilon_X$ and $\mu_X = \sqrt{\frac{x^2}{2 m_X} \ln\left(\frac{2}{\epsilon_X} \right) }$.

Let us begin by developing an intuition for those two different notions of key rates on fluctuating channels. In Figure \ref{fig:varLengthKRs_vs_Runs}, we simulate $100$ runs of the `Statistically Fluctuating Channel' with the following channel parameters $\mu_{\mathrm{sol}} = 2 \times 10^4$, $\sigma_{\mathrm{sol}} = 1\times 10^4$, $\mu_{\mathrm{loss}} = 0.99$ and $\sigma_{\mathrm{loss}} = 0.005$. The pair production rate was set to $0.1$ per time frame, the detection efficiency was assumed to be $\eta_D=0.90$, and the dark-count rate was assumed to be $100$ per second. Furthermore, we set $t_F = 0.5$. We plot the secure key rate for each run, for both the fixed-length and for the variable-length protocol. While the fixed-length protocol either produces a key of fixed length or aborts, it can be seen readily how the key lengths for the variable-length protocol adapt to the observed data, while it only fails for two out of $100$ data points, yielding zero key. In the case of acceptance, the difference in key rate between the fixed-length and the variable-length case is (in this example) about $20\%$. However, since for the fixed-length scenario more than half of the runs are not accepted, this results in the average variable-length key rates being roughly $2.5$ times higher than the fixed-length key rates.

In real scenarios, the channel parameters are usually unknown, so, for the present example, we chose $t_F$ heuristically. This leaves the question of whether, if the behavior of the channel - including its statistical parameters - were known, one could find the optimal $t_F$ for the fixed-length scenario and close the gap in performance. We are going to explore this question in what follows.

In Figure \ref{fig:varLengthKRs_vs_t}, we examine the key rates for the fixed-length protocol and its acceptance ratio for different values of $t_F$. Again, we assumed a `Statistically Fluctuating Channel' with the channel parameters mentioned above and simulated $1000$ runs for each datapoint. The blue dashed line represents the 'one-shot` fixed-length key rates, which are the fixed-length key rates upon acceptance, while the light-red dotted curve shows the associated ratio of accepted rounds. The dot-dashed red curve represents the average fixed-length key rate for each data point (which is the product of the one-shot value with the acceptance ratio). For comparison, we also plot the (average) variable-length key rate (black dotted line). Note that the variable-length key rate does not require $t_F>0$ and thus remains constant with $t_F$. One can see that in this example, the one-shot fixed-length key rates are higher than the variable-length key rates for $t_F < 0.25$. However, due to the low acceptance ratio, the expected fixed-length key rates always remain significantly lower than the variable-length rates. For higher values of $t_F$, the acceptance rate slowly converges toward $1$ while at the same time, the one-shot key rate drops, resulting overall in a decreasing average key rate. The highest fixed-length key rate is observed for $t_F = 1$. However, the exact location of the maximum depends on the channel parameters. Thus, even in case one manages to determine the optimal $t_F$ \textit{a priori}, the obtained average fixed-length key rates would still be significantly below the variable-length rates.

\begin{figure}
\includegraphics[width=0.45\textwidth]{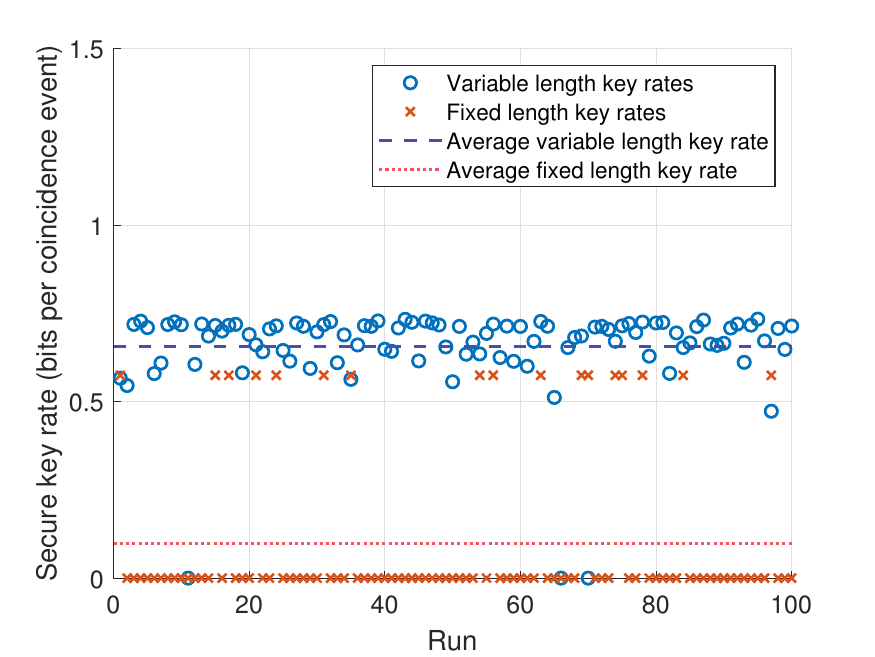}\\
\caption{Comparing variable-length and fixed-length key rates for the `Rapidly Fluctuating Channel'. The system dimension was fixed to $d=16$. \label{fig:comp_chaotic_channel}}
\end{figure}

Already in the case of stable channels with parameters fluctuating statistically around fixed mean values, variable-length key rates outperform their fixed-length equivalents in every aspect. If, in addition, the mean values fluctuate rapidly, the fixed-length acceptance test will only pass in a tiny fraction of all runs, leading to an expected key rate of nearly zero, while the variable-length key adapts to the observed statistics and will produce a nonzero key in many of those cases. We illustrate this in Figure \ref{fig:comp_chaotic_channel}, where we present the results of simulations of the the chaotic channel with parameters $(p_{\textrm{loss}}, n_{\text{sol}}) \in \{ 0.97, 0.98, 0.99 \} \times \{1\times 10^4, 2\times 10^4, 1 \times 10^5\}$ (see Appendix \ref{APDX:SimChannel} for details), where each channel parameter is chosen randomly according to a uniform distribution and $t_F = 0.5$. As expected, the fixed-length protocol aborts in the majority of all cases, yielding zero key. In contrast, the variable-length protocol produces a non-negative key for most runs. The average variable-length rates are now about $6.7$ times higher than the average fixed-length key rate. This illustrates that, in real free-space channels, the advantage of the variable-length security argument will be even more striking.

\section{Discussion}
In recent years, the development of implementations for high-dimensional QKD on the one side and security analyses on the other side has proceeded at a vastly different paces. While highly entangled states have been produced in various degrees of freedom \cite{Kwiat_1997, Barreiro_2005, Martin_2017, Islam_2017, Bulla_2023, Sulimany_2023, Hu_2020, Bouchard_2021, Schneeloch_2019, Ponce_2022, Maxwell_2022}, theoretical analyses have lagged lagging behind, either relying on practically infeasible measurements \cite{Doda_2021, Sheridan_2010} or using computationally intensive convex optimization methods \cite{BKPH_2023} that are limited to very low dimensions, despite taking hours to days to compute. While this problem was solved in the asymptotic regime by Ref. \cite{Kanitschar_2024a}, the practically relevant finite-size regime for HD-QKD was still missing. 

In this work, we close this gap by proving the first composable finite-size security argument for a general high-dimensional QKD protocol against both collective and coherent attacks. The obtained rates converge to the corresponding asymptotic rates  \cite{Kanitschar_2024a} in the limit of large $N$. We obtain non-negative key rates already for practically achievable block sizes of $N=10^8$ under viable conditions. This shows that HD-QKD is suitable for the loss and noise regimes \cite{Liao_2017, Bedington_2017} present in free-space and LEO satellite QKD links. Additionally, we demonstrate that our method provides key rates comparable to those obtained by state-of-the-art numerical convex optimisation techniques. While the latter are computationally intense and thus limited to the low double-digit dimensions even on clusters \cite{BKPH_2023}, our semi-analytic method unlocks genuine high dimensions, providing key rates that surpass existing results in terms of both achievable key rate and maximum tolerable loss.

In addition to common fixed-length key rates, we provide more practically useful variable-length key rates, which build the composability argument around the observed statistics rather than the expected statistics, significantly improving both the practicality and key rates of the protocol. We discuss and compare these two different notions of security to illustrate the advantages of variable-length key rates. 

While we used a time-bin implementation for demonstration purposes, our work is general and can be applied to a variety of platforms with little to no adaptation. In summary, this work provides the missing tool for the full security analysis of general high-dimensional QKD protocols and unlocks the practical use of high-dimensionally entangled states in Quantum Key Distribution. The obvious next steps consist of the application to real implementations and further improvements in the key rate.

\begin{acknowledgements}
F.K. thanks Devashish Tupkary for inspiring discussions and precious feedback on an earlier version of this manuscript,  Ian George for insightful discussions, and Shlok Nahar and Matej Pivoluska for valuable comments on an earlier version of the manuscript. This work has received funding from the Horizon Europe research and innovation programme under grant agreement No 101070168 (HyperSpace). F.K. gratefully acknowledges support from the Dieberger-Skalicky foundation. The authors acknowledge TU Wien Bibliothek for financial support through its Open Access Funding Programme.

\begin{figure}[htb!]
\centering
\includegraphics[width=0.4\columnwidth]{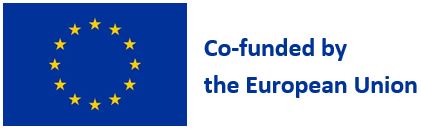}
\end{figure}
 \end{acknowledgements}
 
\appendix
\onecolumngrid
\newpage
\section{Construction of the set $V(\vec{F}^{\mathrm{obs}})$} \label{APDX:Construction_V}
In the fixed-length scenario, the acceptance test was built upon defining expected statistics and finding the set of density matrices that could have produced these statistics only with some small probability; the complement of this set was the set $\mathcal{S}_{\mathrm{AT}}$. In the variable-length scenario, the construction of the acceptance set follows a slightly different logic. A fixed but otherwise unknown quantum state gives rise to observed statistics $\vec{F}^{\mathrm{obs}}$, which is a random variable. Based on this observation, we aim to construct a set $V(\vec{F}^{\mathrm{obs}})$ such that $\mathrm{Pr}\left[\tau_{AB} \in V(\vec{F}^{\mathrm{obs}})\right] \geq 1-\epsilon_{\mathrm{AT}}$. We are using observables, thus we have to slightly adapt the argument from \cite{Tupkary_2024}. Let $\Theta$ denote the set of observables used in the protocol and let $x := ||X||_{\infty} = ||X||_{\infty}||\rho||_1 \geq ||X \rho ||_1$ for $X\in \Theta$. For each $X \in \Theta$, Hoeffding's inequality \cite{Hoeffding_1963} tells us 
\begin{equation}
   \mathrm{Pr}\left[ \left| F_X^{\mathrm{obs}} - \mathbb{E}[X] \right| \geq \mu_X \right] \leq 2 e^{-\frac{2 m_X \mu_X^2}{4 x^2}} =: \epsilon_X, 
\end{equation}
where $m_X$ is the number of rounds used for testing observable $X$. We apply Hoeffding's inequality to every observable $X \in \Theta$ and combine them, using Boole's inequality,
\begin{equation}
   \begin{aligned}
       \mathrm{Pr}\left[\bigcup_{X \in \Theta} \mathbb{E}\left[X\right] \notin \left[F_X^{\mathrm{obs}} - \mu_X, F_X^{\mathrm{obs}} + \mu_X\right] \right]\leq \sum_{X \in \Theta} \epsilon_X = \epsilon_{\mathrm{AT}}. 
   \end{aligned} 
\end{equation}
In reverse, this means
\begin{equation}
    \mathrm{Pr}\left[\bigcap_{X \in \Theta} \mathbb{E}\left[X\right] \in \left[F_X^{\mathrm{obs}} - \mu_X, F_X^{\mathrm{obs}} + \mu_X\right] \right]\geq 1 - \epsilon_{\mathrm{AT}}
\end{equation}
and thus
\begin{equation}
    \mathrm{Pr}\left[ \rho \in V(\vec{F}^{\mathrm{obs}}) \right] \geq 1-\epsilon_{\mathrm{AT}},
\end{equation}
where $\vec{F}^{\mathrm{obs}}$ is defined as the vector containing $F_X^{\mathrm{obs}}$ in it's $X$-th component.

To simplify application, from now on, we require that all the $\epsilon_X$ are equal, hence $\epsilon_{\mathrm{AT}} = |\Theta| \epsilon_X$ and $\mu_X = \sqrt{\frac{x^2}{2 m_X} \ln\left(\frac{2}{\epsilon_X} \right) }$.

\section{Implementation Details} \label{APDX:ImpDetails}
In this section, we give details about implementation and parameter choices used to illustrate our security argument, in particular our choices for the $\epsilon$ scaling for the lift to coherent attacks. After a general discussion, we start with the fixed-length scenario, followed by the more complex variable-length scenario. 

Recall the different security statements in Section \ref{sec:SecurityArgument}, which compose a total security parameter $\epsilon$ from various sub-epsilon parameters such as $\epsilon_{\mathrm{EV}}, \epsilon_{\mathrm{PA}}, \epsilon_{\mathrm{AT}}$ and $\tilde{\epsilon}$. The first two, $\epsilon_{\mathrm{EV}}, \epsilon_{\mathrm{PA}}$, are security parameters related to classical subroutines of the QKD protocol. It is comparably `cheap' to make them almost arbitrarily small, as they scale roughly like $\mathcal{O}\left( 2^{-b} \right)$, where $b$ is the hash length which is the cost in key rate. The third epsilon, $\epsilon_{\mathrm{AT}}$ refers to the acceptance test. As can be seen in Theorem \ref{Thm:AcceptanceTest}, the smaller this parameter, the larger the $\mu$-balls defining the acceptance set. As optimizing over a larger set may lower the key rate significantly, the cost in making $\epsilon_{\mathrm{AT}}$ small is large. Finally, $\tilde{\epsilon}$ is a  'virtual` parameter that does not correspond to any subroutine of the QKD protocol. However, as can be seen by inspecting their corresponding correction terms in Eqs. (\ref{eq:secStatement_eq}) and (\ref{eq:VarLengthCohKRFormula}), their impact on the key rate scales comparably to $\epsilon_{\mathrm{PA}}$ and $\epsilon_{\mathrm{EV}}$. 

We denote the aimed total security parameter for collective attacks by $\epsilon$ and the aimed total security parameter for coherent attacks by $\nu$ to avoid confusion. Keeping the discussed impacts of the different epsilons on the key rate in mind, we aim to maximise the acceptance testing security parameter, at the cost of making all the other epsilons small. To allow for parametrized statements without having to choose numerical values right away, we introduce the parameters $r,s \in [0,1]$.

\subsection{Fixed-length collective attacks}
Recall from Section \ref{sec:SecurityArgument} that the fixed-length $\epsilon$-security statement against collective attacks provides a total security parameter of $\epsilon = \epsilon_{\mathrm{corr}} + \epsilon_{\mathrm{sec}}$, where $\epsilon_{\mathrm{corr}} = \epsilon_{\mathrm{EC}}$ and $\epsilon_{\mathrm{sec}} = \max\{\epsilon_{\mathrm{AT}}, \epsilon_{\mathrm{PA}}\}$. Let $r$ describe how the total security epsilon is split between correctness and secrecy, $\epsilon_{\mathrm{corr}} = r \epsilon$ and consequently $\epsilon_{\mathrm{sec}} = (1-r) \epsilon$. Then, in order to exploit the maximum in the secrecy statement, we choose $\epsilon_{\mathrm{AT}} = \epsilon_{\mathrm{PA}}$. Furthermore, for simplicity, we choose $\epsilon_{\mathrm{PA}} = \bar{\epsilon} = \left(1-r\right) \epsilon$.
This leads to
\begin{align*}
    -2 \log_2\left(\frac{1}{\epsilon_{\mathrm{PA}}}\right) = -2 \log_2\left(\frac{1}{(1-r)\epsilon}\right).
\end{align*}

Finally, for the error-verification term, we simply obtain 
\begin{align*}
    -\frac{1}{N} \log_2\left( \frac{2}{\epsilon_{\mathrm{EV}}}\right) = -\frac{1}{N} \log_2\left( \frac{2}{r \epsilon}\right),
\end{align*}
and
\begin{align*}
    \mu & = \sqrt{\frac{2x^2}{k_T} \ln\left(\frac{2}{\epsilon_{\mathrm{AT}}} \right)} = \sqrt{\frac{2x^2}{k_T} \ln\left(\frac{2}{(1-r)\epsilon} \right)}.
\end{align*}

\subsection{Fixed-length coherent attacks}
We performed a similar strategy for the coherent attack scenario. To ensure comparability, we again aim for a total (adjusted) security parameter of $\nu = 10^{-10}$. However, due to the corrections listed in Theorem \ref{thm:SecurityStatementCoh}, $\nu_{\mathrm{corr}} =\epsilon_{\mathrm{EV}}$, $\nu_{\mathrm{sec}} =  g_{n,x} \left( \epsilon_{\mathrm{PA}} + 2 \sqrt{2 \epsilon_{\mathrm{AT}}}\right)$, we follow a slightly different strategy. Let us discuss this in more detail and denote the scaled security parameters of the subprotocols by $\nu$. While $\nu_{\mathrm{EV}} := \epsilon_{\mathrm{EV}} = r \nu$ remains untouched and we still have $\nu_{\mathrm{sec}} = (1-r) \nu$, let us define $\nu_{\mathrm{PA}} := g_{n,x} \epsilon_{\mathrm{PA}}$, and $\nu_{\mathrm{AT}} := 2 g_{n,x} \sqrt{2\epsilon_{\mathrm{AT}}}$. Now we make use of the other parameter $s$ to describe the splitting between the different contributions in the secrecy epsilon. Let $\nu_{\mathrm{PA}} = s \nu_{\mathrm{sec}} = (1-r)s \nu$ and $\nu_{\mathrm{AT}} = (1-s) \nu_{\mathrm{sec}} = (1-r)(1-s) \nu$. As discussed already before, this was motivated by the following observation. Despite the apparently massive correction factor of $g_{n,x} = {n+x-1 \choose n} $, since $\epsilon_{\mathrm{PA}}$ scales inverse exponentially with the number of hash bits, it can be compensated for a very moderate price of shortening the key by only $\log_2\left( g_{n,x} \right)$,
\begin{align*}
    -\frac{2}{N} \log_2\left( \frac{1}{\epsilon_{\mathrm{PA}}}\right) \mapsto& -\frac{2}{N} \log_2\left( \frac{g_{n,x}}{ \nu_{\mathrm{PA}}}\right) =-\frac{2}{N} \log_2\left( \frac{1}{ (1-r)s \nu}\right) -\frac{2}{N} \log_2\left( g_{n,x}\right)
\end{align*}

However, while we can compensate the correction for $\epsilon_{\mathrm{AT}}$ as well, it increases $\mu_X$, hence the feasible set, which affects the key rate significantly,
\begin{align*}
    \mu & = \sqrt{\frac{2x^2}{k_T} \ln\left(\frac{2}{\epsilon_{\mathrm{AT}}} \right)} \mapsto \mu' = \sqrt{\frac{2x^2}{k_T} \ln\left(\frac{16 g_{n,x}^2}{\nu_{\mathrm{AT}}^2} \right)} =  
    \sqrt{\frac{2x^2}{k_T} \left( \log_2\left(\frac{16}{(1-r)^2(1-s)^2 \nu^2} \right) + 2 \log_2 \left( g_{n,x}\right) \right)}.
\end{align*}

Finally, for the error-verification term, we simply obtain 
\begin{align*}
    -\frac{1}{N} \log_2\left( \frac{2}{\epsilon_{\mathrm{EV}}}\right) \mapsto -\frac{1}{N} \log_2\left( \frac{2}{r \nu_{\mathrm{EV}}}\right)
\end{align*}

\subsection{Variable-length collective attacks}
This leads us to the variable-length scenario. Again, we start with the collective attacks security statement, where $\epsilon_{\mathrm{corr}} = \epsilon_{\mathrm{EV}} = r \epsilon$ and $\epsilon_{\mathrm{sec}} = \epsilon_{\mathrm{PA}} + \epsilon_{\mathrm{AT}} = (1-r) \epsilon$. Similar to before, we set $\epsilon_{\mathrm{PA}} = s \epsilon_{\mathrm{sec}} = (1-r)s \epsilon$ and consequently $\epsilon_{\mathrm{AT}} = (1-s) \epsilon_{\mathrm{sec}} = (1-r)(1-s) \epsilon$. Then, we obtain
\begin{align*}
    \alpha = 1 + \frac{\sqrt{\log_2\left(\frac{1}{\epsilon_{\mathrm{PA}}} \right)}}{\log_2(d+1)\sqrt{n}} = 1 + \frac{\sqrt{\log_2\left(\frac{1}{r\epsilon} \right)}}{\log_2(d+1)\sqrt{n}},
\end{align*}
\begin{align*}
    \theta = \frac{\alpha}{\alpha-1}\left( \log_2\left( \frac{1}{4\epsilon_{\mathrm{PA}}} + \frac{2}{\alpha}\right) \right) + \left\lceil \log_2\left(\frac{2}{\epsilon_{\mathrm{EV}}}\right) \right\rceil = \frac{\alpha}{\alpha-1}\left( \log_2\left( \frac{1}{4(1-r)s \epsilon} + \frac{2}{\alpha}\right) \right) + \left\lceil \log_2\left(\frac{2}{r\epsilon}\right) \right\rceil,
\end{align*}
and
\begin{align*}
    \mu & = \sqrt{\frac{2x}{k_T} \ln\left(\frac{2}{\epsilon_{\mathrm{AT}}} \right)} = \sqrt{\frac{2x}{k_T} \ln\left(\frac{2}{(1-r)(1-s)\epsilon} \right)}.
\end{align*}

\subsection{Variable-length coherent attacks}
Finally, we come to the variable-length coherent attack scenario. Here, according to Theorem \ref{thm:SecurityStatementVariableCoh}, the scaled security statement consists of $\nu_{\mathrm{corr}} = \epsilon_{\mathrm{EV}}$ and $\nu_{\mathrm{sec}} = g_{n,x} \left(\sqrt{8 \left(\epsilon_{\mathrm{PA}} + \epsilon_{\mathrm{AT}}\right)} + \frac{\tilde{\epsilon}}{2} \right)$. Again, we start with $\nu_{\mathrm{corr}} = r \nu$ and $\nu_{\mathrm{sec}} = (1-r) \nu$, which we are going to use below.

Next, we set $g_{n,x} \frac{\tilde{\epsilon}}{2} = s \nu_{\mathrm{sec}} = (1-r)s \nu$, which yields
\begin{align*}
    -2 \log_2\left(\frac{1}{\tilde{\epsilon}}\right) \mapsto -2 \log_2\left(\frac{2(1-r)s}{\nu} \right) - 2 \log_2\left( g_{n,x}\right)
\end{align*}
and leaves us back with $g_{n,x} \sqrt{8} \sqrt{\epsilon_{\mathrm{PA}} + \epsilon_{\mathrm{AT}}} = (1-s) \nu_{\mathrm{sec}} = (1-r)(1-s)\nu$. Additionally, we set $\epsilon_{\mathrm{PA}} = s\frac{(1-s)^2 \nu_{\mathrm{sec}}^2}{8 g_{n,x}^2} = s\frac{(1-s)^2 (1-r)^2 \nu^2}{8 g_{n,x}^2} $ and consequently $\epsilon_{\mathrm{AT}} = (1-s)\frac{(1-s)^2 \nu_{\mathrm{sec}}^2}{8 g_{n,x}^2} = \frac{(1-r)^2(1-s)^3 \nu^2}{8 g_{n,x}^2}$. This now allows us to discuss the first term of $\theta$,
\begin{align*}
    \theta = \frac{\alpha}{\alpha-1}\left[ \log_2\left( \frac{1}{4\epsilon_{\mathrm{PA}}} + \frac{2}{\alpha}\right) \right] &+\left\lceil\log_2\left( \frac{2}{\epsilon_{\mathrm{EC}}} \right)\right\rceil\\ \mapsto \theta' :=& \frac{\alpha}{\alpha-1} \left[ \log_2\left( \frac{8 g_{n,x}^2}{4s(1-s)^2\nu_{\mathrm{sec}}^2} + \frac{2}{\alpha}\right) \right] + \left\lceil\log_2\left( \frac{2}{r \nu} \right)\right\rceil \\
    =& \frac{\alpha}{\alpha-1} \left[ \log_2\left( \frac{2 g_{n,x}^2}{(1-r)^2s(1-s)^2\nu^2} + \frac{2}{\alpha}\right) \right] +\left\lceil\log_2\left( \frac{2}{r \nu} \right)\right\rceil\\
    =&\frac{\alpha}{\alpha-1} \left[ \log_2\left( g_{n,x}^2\left( \frac{2}{(1-r)^2s(1-s)^2\nu^2} + \frac{2}{\alpha g_{n,x}^2}\right)\right) \right] +\left\lceil\log_2\left( \frac{2}{r \nu} \right)\right\rceil\\
    =&\frac{\alpha}{\alpha-1} \left[2 \log_2\left( g_{n,x}\right) + \log_2\left(\left( \frac{2}{(1-r)^2s(1-s)^2\nu^2} + \frac{2}{\alpha g_{n,x}^2}\right)\right) \right] +\left\lceil\log_2\left( \frac{2}{r \nu} \right)\right\rceil\\
    \leq& \frac{\alpha}{\alpha-1} \left[2 \log_2\left( g_{n,x}\right) + \log_2\left(\left( \frac{2}{(1-r)^2s(1-s)^2\nu^2} + \frac{2}{\alpha}\right)\right) \right] +\left\lceil\log_2\left( \frac{2}{r \nu} \right)\right\rceil.
\end{align*}

It remains to discuss the scaling of $\mu$,
\begin{align*}
    \mu = \sqrt{\frac{2x}{k_T} \ln\left(\frac{2}{\epsilon_{\mathrm{AT}}} \right)} \mapsto \sqrt{\frac{2x}{k_T} \ln\left(\frac{16 g_{n,x}^2}{(1-s)^3 \nu_{\mathrm{sec}}^2} \right)} = \sqrt{\frac{2x}{k_T} \left[ \ln\left(\frac{16}{(1-r)^2(1-s)^3 \nu^2} \right) + 2 \ln\left(g_{n,x}\right) \right]}.
\end{align*}

Note that we may choose $\alpha$ freely, so in particular according to the security parameters after the lift. Thus, we have
\begin{align*}
    \alpha = 1 + \frac{\sqrt{\log_2\left(\frac{1}{\epsilon_{\mathrm{PA}}} \right)}}{\log_2(d+1)\sqrt{n}} = 1 + \frac{\sqrt{\log_2\left(\frac{1}{r\epsilon} \right)}}{\log_2(d+1)\sqrt{n}},
\end{align*}

\section{Simulation of the Atmospheric Channel}\label{APDX:SimChannel}
In order to demonstrate the advantages of a variable-length security argument, we simulate secure key rates for two different channel behaviors. First, we consider a statistically fluctuating channel, followed by a rapidly fluctuating channel. Both simulations use the noise model from \cite{BKPH_2023} to relate the experimental quantities pair-production rate $c_{\mathrm{Prod}}$, dark-count rate $d_{\mathrm{dark}}$, detection efficiency $\eta_d$, channel loss probability $p_L$ and number of environmental (solar) photons $n_{\mathrm{sol}}$ to click matrices. For details regarding this noise model, we refer interested readers to Ref. \cite{BKPH_2023}.

\subsection{Statistically Fluctuating Channel}
For this scenario, we assume that both the setup and the channel behavior are relatively well-known in advance. The system is characterized by a fixed pair-production rate $c_{\mathrm{Prod}}$, a fixed dark-count rate $d_{\mathrm{dark}}$, and a fixed detection efficiency $\eta_d$. The channel loss and the number of environmental photons entering the detection device are assumed to follow a normal distribution with means $\mu_{\mathrm{loss}}$ and $\mu_{\mathrm{sol}}$ and standard deviations $\sigma_{\mathrm{loss}}$ and $\sigma_{\mathrm{sol}}$. For our simulation we then sample $n_{\mathrm{sim}}$-times $p_l ~ \mathcal{N}(\mu_{\mathrm{loss}}, \sigma_{\mathrm{loss}})$ and $n_{\mathrm{sol}} ~\mathcal{N}(\mu_{\mathrm{sol}},\sigma_{\mathrm{sol}})$ while keeping all other parameters fixed. In the fixed-length scenario, within the frame of the acceptance test, the observables ($\hat{W}_1$ and $\hat{W}_2$) are compared to the values $w_1^U$ and $w_2$. In case they lie within a $t$-box around those values, the protocol accepts, and we obtain a key of length $l$. Otherwise, it aborts (and we do not obtain the key). In the variable-length scenario, each of those runs may potentially yield a secure key, albeit the length differs and depends on the observed values.

\subsection{Rapidly Fluctuating Channel}
For this scenario, we proceed assuming that the system is well-characterized. This is the pair-production rate $c_{\mathrm{Prod}}$, the dark-count rate $d_{\mathrm{dark}}$, and the detection efficiency $\eta_d$ remain fixed and constant. At the same time, now we do not know much about the channel. Thus neither the loss nor the number of environmental photons are known in advance, as they change rapidly. To ease simulation, we nevertheless assume that the conditions only rapidly change between a discrete number of scenarios. In more detail, we model $3$ brightness levels ('pitch dark`, 'dark` and 'dawn`) as well as $3$ loss levels ('no clouds`, 'slightly cloudy` and 'cloudy`) leading to a total of $9$ combined scenarios, described by pairs of parameters 7. While, if known as honest channel behavior is known in advance, in each of those scenarios one could generate secure keys, in case of rapid changes, e.g., if a cloud moves into the direct line of sight between transmitter and receiver, the fixed length protocol aborts in $8$ out of those $9$ cases, failing to produce a key. This is because the acceptance test aborts. For simplicity, we assume each of those $9$ scenarios occurs with the same probability.

In contrast, the variable-length protocol attempts to generate keys in any of those $9$ scenarios and adapts the secure key length to the measured results.

\twocolumngrid
\bibliographystyle{apsrev4-2}
\bibliography{Bibliography}
	
\end{document}